\documentclass[utf8]{aa}
\pdfoutput=1
\usepackage{graphicx}
\usepackage[varg]{txfonts}
\usepackage{hyperref}
\hypersetup{
        colorlinks   = true, 
        urlcolor     = blue, 
        linkcolor    = blue, 
        citecolor   = blue 
}
\usepackage{subcaption}
\defcitealias{Henriques2015}{H15}
\usepackage{physics}

\usepackage{algpseudocode}
\usepackage{algorithm}  
\usepackage[alsoload=astro]{siunitx}
\allowdisplaybreaks

\begin{document} 

\bibpunct[; ]{(}{)}{;}{a}{}{,}

\algrenewcommand\algorithmicwhile{\textbf{In each thread}}
\algrenewtext{EndWhile}{\algorithmicend\ \textbf{thread}}

\title{Measuring galaxy-galaxy-galaxy-lensing with higher precision and accuracy}
\titlerunning{Galaxy-galaxy-galaxy-lensing with higher precision and accuracy}

\author{Laila Linke \inst{1}
          \and
          Patrick Simon \inst{1}
          \and
          Peter Schneider \inst{1}
          \and
          Stefan Hilbert \inst{2,3}
                }

   \institute{Argelander-Institut f\"ur Astronomie, Universit\"at Bonn, Auf dem H\"ugel 71, 53121 Bonn, Germany\\
   \email{llinke@astro.uni-bonn.de}
   \and 
   Excellenzcluster Universe, Boltzmannstr. 2, 85748 Garching, Germany
   \and
   Ludwig-Maximilians-Universität, Universitäts-Sternwarte,  Scheinerstr.1, 81679 M\"unchen, Germany        }

   \date{Received XXX; accepted YYY}

 \abstract
  {Galaxy-galaxy-galaxy lensing (G3L) is a powerful tool for constraining the three-point correlation between {the galaxy and matter distribution} and thereby models of galaxy evolution.}
   {We propose three improvements to current measurements of G3L: (i) a weighting of lens galaxies according to their redshift difference, (ii) adaptive binning of the three-point correlation function, and (iii) accounting for the effect of lens magnification by the cosmic large-scale structure. Improvement (i) is designed to improve the precision of the G3L measurement, whereas improvements (ii) and (iii) remove biases of the estimator. We further show how the G3L signal  can be converted from angular into physical scales.}
  {The improvements were tested on simple mock data and simulated data based on the Millennium Run with an implemented semi-analytic galaxy model.}
  {Our improvements increase the signal-to-noise ratio by \SI{35}{\percent} on average at angular scales between $\ang[astroang]{;0.1;}$ and $\ang{;10;}$ and physical scales between $0.02$ and $2 \, h^{-1}\, \textrm{Mpc}$. They also remove the bias of the G3L estimator at angular scales below $\ang{;1;}$, which was originally up to $\SI{40}{\percent}$. The signal due to lens magnification is approximately $\SI{10}{\percent}$ of the total signal.}
   {}


   \maketitle
%

\section{Introduction}

In the current standard model of cosmology the majority of matter in the Universe is dark and only interacts gravitationally \citep{Planck2015_CosmicParams,Hildebrandt2017}. While the formation of dark matter halos and the cosmic large-scale structure (LSS) has been successfully modelled by $N$-body simulations \citep[see e.g.][]{Springel2005}, the interplay between dark and baryonic matter is still not well understood. Observational tools are needed to distinguish between various semi-analytic models of galaxy evolution \citep[SAMs; see e.g.][]{Henriques2015,Lacey2016} and to test the predictions of hydrodynamical simulations \citep[see e.g.][]{Eagle2015,Illustris2014}.

One promising tool is galaxy-galaxy-galaxy-lensing (G3L), first proposed by \citet{Schneider2005}. It involves measuring the connected three-point correlation function between the galaxy and matter distribution by either evaluating the gravitational lensing shear of background galaxies around foreground galaxy pairs (lens-lens-shear correlation) or the lensing shear of background galaxy pairs around single foreground galaxies (lens-shear-shear correlation). 

Here, we concentrate on the lens-lens-shear correlation function on small sub-megaparsec scales, where it is most sensitive to galaxy pairs residing in the same matter halo. This function is a powerful discriminator between models of galaxy formation and evolution. In particular, \citet{Saghiha2017} showed that it can better distinguish between galaxy evolution models than galaxy-galaxy-lensing (GGL), where the average shear of {individual source galaxies around individual lens galaxies} is measured \citep[e.g.][]{Mandelbaum2006}

The G3L correlation function on small scales was measured successfully by \citet{Simon2008} in the Red Cluster Sequence Lensing Survey and by \citet{Simon2013} in the Canada-France-Hawaii Telescope Lensing Survey (CFHTLenS). However, these measurements were based on photometric data without precise redshift estimates. Consequently, pairs of lens galaxies that are physically close and therefore highly correlated were treated with the same weight as galaxy pairs that are separated along the line of sight and have little to no correlation. As discussed by \citet{Simon2019}, these separated galaxies decrease the signal and lower the signal-to-noise ratio (S/N).

Related measurements of the correlation of galaxy pairs and the matter distribution were also undertaken on larger scales, with galaxy pairs separated by several megaparsec, to detect inter-cluster filaments \citep{Mead2010, Clampitt2016, Epps2017, Kondo2019, Xia2019}. These studies relied on precise galaxy redshift estimates provided by spectroscopic surveys. This paper investigates how similarly precise redshift information can be used to enhance the S/N of G3L at smaller scales.

Additionally, G3L is affected by the magnification of lens galaxies caused by the LSS in front of the lenses \citep{Bartelmann2001}. This magnification affects the selection function and thereby the number density of lens galaxies in a survey. Because source galaxies are also lensed by the LSS, the shear of sources is correlated with the lens magnification, and an additional correlation signal arises. This signal has not yet been quantified for G3L, but was found to affect GGL by up to \SI{5}{\percent} in CFHTLenS \citep{Simon2018}.

We introduce three improvements to the G3L estimator used by \citet{Simon2008, Simon2013}: (i) weighting the lens galaxy pairs according to their redshift difference, (ii) using a new, adaptive binning method for the correlation function to reduce biases, and (iii) estimating the magnification bias with lens galaxies that are separated along the line of sight. We also show how the correlation can be measured in terms of physical instead of angular separation and weight the signal by the critical surface mass density $\Sigma_\text{crit}$, as is common for GGL \citep[e.g.][]{Mandelbaum2006}. Thereby, the signal no longer depends on the redshift distribution of source galaxies. To test the effect of our improvements, we apply the new estimator to simple mock data, for which we can directly calculate the expected aperture statistics, and to simulated data based on the Millennium Run \citep[MR]{Springel2005} with the SAM by \citet[H15]{Henriques2015}.

This paper is structured as follows: Section~\ref{sec:fundamentals} defines the fundamental quantities of G3L, and gives the estimator for the three-point correlation function by \citet{Simon2008}. Section~\ref{sec:methods} explains our new estimator with redshift weighting and the new binning scheme, as well as how the estimator can be converted into physical units and the effect of lens magnification can be estimated. We describe our simulated data set from the MR in Sect.~\ref{sec:data}. The results of applying our improved measurement scheme to the data are given in Sect.~\ref{sec:results}, and they are discussed in Sect.~\ref{sec:discussion}


\section{Fundamentals of galaxy-galaxy-galaxy-lensing}
\label{sec:fundamentals}

\begin{figure}
\centering
\resizebox{\hsize}{!}{\includegraphics{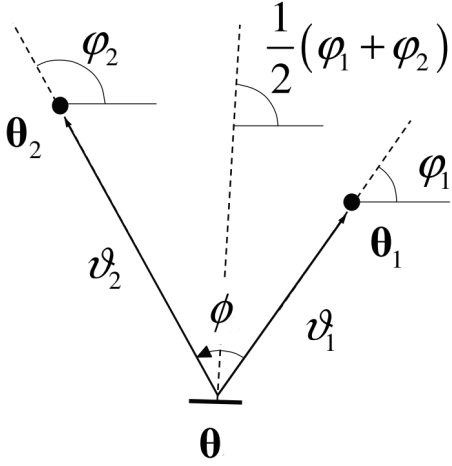}}
\caption{Geometry of a G3L configuration with one source and two lens galaxies; adapted from \citet{Schneider2005}.}
\label{fig:G3L}
\end{figure}

The G3L is a weak gravitational lensing effect, first discussed by \citet[for a review on weak lensing, see  \citealp{Bartelmann2001}]{Schneider2005}.  We concentrate on systems of two lens and one source galaxy, whose geometric configuration projected onto the sky is shown in Fig.~\ref{fig:G3L}. The main observable of G3L in these systems is the three-point correlation function $\tilde{\mathcal{G}}$ of the projected lens galaxy number density $N(\va*{\vartheta})$ and tangential gravitational lensing shear $\gamma_\textrm{t}$, given by
\begin{equation}
\label{eq:Gtilde}
\tilde{\mathcal{G}}({\va*{\vartheta}_1}, {\va*{\vartheta}_2})=\frac{1}{\overline{N}^2}\,\expval{ N(\va*{\theta}+\va*{\vartheta}_1)\, N(\va*{\theta}+\va*{\vartheta}_2) \, \gamma_\textrm{t}(\va*{\theta})}\;.
\end{equation}
The tangential shear is measured with respect to the bisector of the angle $\phi$ between the lens positions $\va*{\theta} + \va*{\vartheta}_1$ and $\va*{\theta} + \va*{\vartheta}_2$. Because of the isotropy and homogeneity of the matter density field, $\tilde{\mathcal{G}}$ only depends on the lens-source separations $\vartheta_1$ and $\vartheta_2$ and on the opening angle $\phi$, so we write
\begin{equation}
\label{eq:GtildeNewNotation}
\tilde{\mathcal{G}}(\va*{\vartheta}_1, \va*{\vartheta}_2) \equiv \tilde{\mathcal{G}}(\vartheta_1, \vartheta_2, \phi)\;.
\end{equation}

We can estimate $\tilde{\mathcal{G}}(\vartheta_1, \vartheta_2, \phi)$ by averaging the tangential ellipticities of all lens-lens-source triplets where $\va*{\vartheta}_1\,(\va*{\vartheta}_2)$ is the separation between the first (second) lens and the source. As discussed by \citet{Simon2008}, this average is an estimator of 
\begin{equation}
\label{eq:sum_estimator}
\frac{\expval{N(\va*{\theta} + \va*{\vartheta}_1)\, N(\va*{\theta} +\va*{\vartheta}_2)\, \gamma_t(\va*{\theta})}}{\expval{N(\va*{\theta} + \va*{\vartheta}_1) \,N(\va*{\theta} + \va*{\vartheta}_2)}} = \frac{\tilde{\mathcal{G}}(\vartheta_1, \vartheta_2, \phi)}{1+\omega(|\va*{\vartheta}_1 - \va*{\vartheta}_2|)}\,,
\end{equation}
with the angular two-point correlation function $\omega$ of lens galaxies. When the complex ellipticity $\epsilon_k$ of source galaxies is used as estimator of their complex lensing shears, Eq.~\eqref{eq:sum_estimator} implies that $\tilde{\mathcal{G}}$ can be estimated for each bin $b$ of $\vartheta_1$, $\vartheta_2$ and $\phi$ by a triple sum over all $N_\textrm{d}$ lenses and $N_\textrm{s}$ sources by 
\begin{align}
\label{eq:Gtilde_est}
&\notag \tilde{\mathcal{G}}_\text{est}(b) \\
&= -\dfrac{\sum_{i,j=1}^{N_\textrm{d}}\sum_{k=1}^{N_\textrm{s}} w_k \, \epsilon_k \,  \textrm{e}^{-\textrm{i}(\varphi_{ik}+\varphi_{jk})} \, \left[1+\omega(|\va*{\theta}_i -\va*{\theta}_j|)\right] \,  { \Delta}_{ijk}(b)}{\sum_{i,j=1}^{N_\textrm{d}}\sum_{k=1}^{N_\textrm{s}} \, w_k \, \Delta_{ijk}(b)}\\
&=: -\dfrac{\sum_{ijk} w_k \, \epsilon_k \,  \textrm{e}^{-\textrm{i}(\varphi_{ik}+\varphi_{jk})} \, \left[1+\omega(|\va*{\theta}_i -\va*{\theta}_j|)\right] \,  \Delta_{ijk}(b)}{\sum_{ijk} \, w_k \, \Delta_{ijk}(b)}\; ,
\end{align}
with
\begin{equation}
\label{eq:vartriangle}
\Delta_{ijk}(b) = 
\begin{cases}
1 &\textrm{for}\left(|\va*{\theta}_k-\va*{\theta}_i|,|\va*{\theta}_k-\va*{\theta}_j|,\phi_{ijk}\right) \in b \\
0&\textrm{otherwise}
\end{cases}\;.
\end{equation}
The angle $\varphi_{ik}$($\varphi_{jk}$) is the polar angle of the lens-source separation vector $\va*{\theta}_i-\va*{\theta}_k$ ($\va*{\theta}_j-\va*{\theta}_k$), and $\phi_{ijk}$ is the opening angle between $\va*{\theta}_i-\va*{\theta}_k$ and $\va*{\theta}_j-\va*{\theta}_k$ (see Fig.~\ref{fig:G3L}; $\va*{\theta}_i-\va*{\theta}_k$ and $\va*{\theta}_j-\va*{\theta}_k$ correspond to $\va*{\vartheta}_1$ and $\va*{\vartheta}_2$, $\phi_{ijk}$ corresponds to $\phi$). The $w_k$ are weights of the measured ellipticities. To apply the estimator to observational data, a higher weight should be assigned to sources with more precise shape measurements, whereas sources with less precise shapes should be down-weighted. Because we apply the estimator to simulated data, we set $w_k \equiv 1$ throughout this work. The phase factor and minus sign in Eq.~\eqref{eq:Gtilde_est} are due to the definition of tangential and cross shear in terms of the Cartesian shear components.

In order to estimate $\omega$, we use ``randoms''. These are mock galaxies that are distributed unclustered on the sky, but obey the same selection function as the lens galaxies. With these randoms, we estimate $\omega$ with the Landy--Szalay estimator \citep{LandySzalay1993}
\begin{equation}
\label{eq:LandySzalay}
\omega(\theta) = \frac{N_\textrm{r}^2 \, DD(\theta)}{N_\textrm{d}^2 \, RR(\theta)} - 2 \,\frac{N_\textrm{r} \, DR(\theta)}{N_\textrm{d}\, RR(\theta)} + 1\;.
\end{equation}
Here, $DD(\theta)$ is the pair-count of the $N_\textrm{d}$ lens galaxies, $RR(\theta)$ is the pair-count of the $N_\textrm{r}$ randoms and $DR(\theta)$ is the cross pair-count of lenses and randoms at separation $\theta$.

The correlation function $\tilde{\mathcal{G}}$ mixes second- and third-order statistics. This becomes evident when we rewrite $\tilde{\mathcal{G}}$, using the galaxy density contrast $\kappa_\textrm{g}={N}/{\bar{N}}-1$, as
        \begin{align}
        \label{eq:GtildeWithGGL}
                \tilde{\mathcal{G}}(\va*{\vartheta}_1, \va*{\vartheta}_2) &= \expval{\kappa_\textrm{g}(\va*{\theta}+\va*{\vartheta}_1)\, \kappa_\textrm{g}(\va*{\theta}+\va*{\vartheta}_2)\, \gamma_\textrm{t}(\va*{\theta})}\\
                &\notag \quad + \expval{\kappa_\textrm{g}(\va*{\theta}+\va*{\vartheta}_1)\, \gamma_\textrm{t}(\va*{\theta})} + \expval{\kappa_\textrm{g}(\va*{\theta}+\va*{\vartheta}_2)\, \gamma_\textrm{t}(\va*{\theta})}\\
                &=: \mathcal{G}(\va*{\vartheta}_1, \va*{\vartheta}_2)\\
                &\notag \quad + \expval{\kappa_\textrm{g}(\va*{\theta}+\va*{\vartheta}_1)\, \gamma_\textrm{t}(\va*{\theta})} + \expval{\kappa_\textrm{g}(\va*{\theta}+\va*{\vartheta}_2)\, \gamma_\textrm{t}(\va*{\theta})}\;.
        \end{align}
The second and third term in Eq.~\eqref{eq:GtildeWithGGL} are the GGL signals around individual lenses, while only the first term encompasses the G3L signal, which is the additional correlation $\mathcal{G}$ around lens pairs.

To remove the contribution due to GGL, we convert $\tilde{\mathcal{G}}$ into aperture statistics. Aperture statistics are expectation values of products of the aperture number count $\mathcal{N}_\theta$ and the aperture mass $M_{\rm ap, \theta}$. These are defined as \citep{Bartelmann2001}
\begin{equation}
\label{eq:DefinitionApertureNumberCount}
\mathcal{N}_\theta(\va*{\vartheta}) = \frac{1}{\overline{N}}\int \dd[2]{\vartheta'} \,U_\theta(|\va*{\vartheta}-\va*{\vartheta}'|)\, N(\va*{\vartheta}')\; ,
\end{equation}
and
\begin{equation}
\label{eq:DefinitionApertureMass}
{M}_{\rm ap, \theta}(\va*{\vartheta}) = \int \dd[2]{\vartheta'} \,U_\theta(|\va*{\vartheta}-\va*{\vartheta}'|)\, \kappa(\va*{\vartheta}')\; ,
\end{equation}
with the projected galaxy number density $N(\va*{\vartheta})$, the lensing convergence $\kappa(\va*{\vartheta})$, and the filter function $U_\theta(\vartheta)$ with characteristic scale $\theta$. This filter function needs to be compensated for, that is, $\int_0^{\infty} \dd{\vartheta} \vartheta \,U_{\theta}(\vartheta) = 0$. Because of this property, the aperture number count can be written in terms of the galaxy number density contrast as
\begin{equation}
\label{eq:DefinitionApertureNumberCount_kappa}
\mathcal{N}_\theta(\va*{\vartheta}) = \int \dd[2]{\vartheta'} \,U_\theta(|\va*{\vartheta}-\va*{\vartheta}'|)\, \kappa_\textrm{g}(\va*{\vartheta}')\; .
\end{equation}
For each $U_\theta$, an associated filter function $Q_{\theta}$ can be defined by
\begin{equation}
\label{eq:FilterQ}
        Q_\theta(\vartheta)= \frac{2}{\vartheta^2}\int_0^{\vartheta}\dd{\vartheta'}\, \vartheta'\, U_\theta(\vartheta') - U_\theta(\vartheta)\;.
\end{equation}
With this filter $Q_\theta$, 
\begin{align}
\label{eq:ApertureMassFromGamma}
&{M}_{\rm ap, \theta}(\va*{\vartheta}) + \textrm{i}\, M_{\perp, \theta}(\va*{\vartheta})\\
&\notag= \int \dd[2]{\vartheta'} \,Q_\theta(|\va*{\vartheta}-\va*{\vartheta}'|)\, \left[\gamma_\textrm{t}(\va*{\vartheta}') + \textrm{i}\,\gamma_\times(\va*{\vartheta}')\right]\;,
\end{align}
where $M_{\perp, \theta}$ is the B mode of the aperture mass.

With the lens-lens-shear correlation function, we study the aperture statistics $\expval{\mathcal{N}^2 M_\textrm{ap}}$ and $\expval{\mathcal{N}^2 M_\perp}$, given by
\begin{align}
\label{eq:Definition NNMap}
&\notag \expval{\mathcal{N}^2 M_\textrm{ap}} (\theta_1, \theta_2, \theta_3) + \textrm{i} \expval{\mathcal{N}^2 M_\perp}(\theta_1, \theta_2, \theta_3) \\
&= \frac{1}{\overline{N}^2}\int \dd[2]{\vartheta_1}\, \int \dd[2]{\vartheta_2} \,\int \dd[2]{\vartheta_3}\, U_{\theta_1}( \vartheta_1) \,U_{\theta_2}({\vartheta_2})\, Q_{\theta_3}(\vartheta_3)\,\\
&\notag\quad \times \expval{N(\va*{\vartheta}_1)\, N(\va*{\vartheta}_2)\, \left[\gamma_\textrm{t}(\va*{\vartheta}_3)+\textrm{i}\, \gamma_\times(\va*{\vartheta}_3)\right]}\,.
\end{align}
These aperture statistics can be related to $\tilde{\mathcal{G}}$ for a chosen filter function $U_\theta$. Provided the exponential filter function,
\begin{equation}
\label{eq:exponentialFilterFunction}
U_\theta(\vartheta) = \frac{1}{2\pi \theta^2} \, \left(1 - \frac{\vartheta^2}{2 \theta^2} \right) \, \exp( -\frac{\vartheta^2}{2 \theta^2} )\;,
\end{equation}
\citet{Schneider2005} found
\begin{align}
\label{eq:NNMapFromG}
&\notag\expval{\mathcal{N}^2 M_\textrm{ap}} (\theta_1, \theta_2, \theta_3)\\
&= \int_0^{\infty}\dd{\vartheta_1}\, \vartheta_1\, \int_0^{\infty}\dd{\vartheta_2}\, \vartheta_2\, \int_0^{2\pi} \dd{\phi}\;\tilde{\mathcal{G}}(\vartheta_1, \vartheta_2, \phi)\\
&\quad \notag \times {A}_{\mathcal{N}\mathcal{N} M}(\vartheta_1, \vartheta_2,\phi\; |\; \theta_1, \theta_2, \theta_3)\;,
\end{align}
with the kernel function ${A}_{\mathcal{N}\mathcal{N} M}(\vartheta_1, \vartheta_2,\phi\; |\; \theta_1, \theta_2, \theta_3)$ in the appendix of \citet{Schneider2005}. 

Due to Eq.~\eqref{eq:DefinitionApertureNumberCount_kappa}, Eq.~\eqref{eq:Definition NNMap} can also be written as
        \begin{align}
        \label{eq:Definition NNMap_kappa}
        &\notag \expval{\mathcal{N}^2 M_\textrm{ap}} (\theta_1, \theta_2, \theta_3) + \textrm{i} \expval{\mathcal{N}^2 M_\perp}(\theta_1, \theta_2, \theta_3) \\
        &= \int \dd[2]{\vartheta_1}\, \int \dd[2]{\vartheta_2} \,\int \dd[2]{\vartheta_3}\, U_{\theta_1}\left( \vartheta_1 \right) \,U_{\theta_2}\left( {\vartheta_2}\right)\, Q_{\theta_3}\left(\vartheta_3\right)\,\\
        &\notag \quad \times \expval{\kappa_\textrm{g}(\va*{\vartheta}_1)\, \kappa_\textrm{g}(\va*{\vartheta}_2)\, \left[\gamma_\textrm{t}(\va*{\vartheta}_3)+\textrm{i}\, \gamma_\times(\va*{\vartheta}_3)\right]}\;,
        \end{align}
which leads to
        \begin{align}
        \label{eq:NNMapFromGnotilede}
&\notag \expval{\mathcal{N}^2 M_\textrm{ap}} (\theta_1, \theta_2, \theta_3)\\
&= \int_0^{\infty}\dd{\vartheta_1}\, \vartheta_1\, \int_0^{\infty}\dd{\vartheta_2}\, \vartheta_2\, \int_0^{2\pi} \dd{\phi}\;\mathcal{G}(\vartheta_1, \vartheta_2, \phi)\\
&\quad \notag \times {A}_{\mathcal{N}\mathcal{N} M}(\vartheta_1, \vartheta_2,\phi\; |\; \theta_1, \theta_2, \theta_3)\;,
        \end{align}
        with the same kernel function as Eq.~\eqref{eq:NNMapFromG}. Consequently, the aperture statistics depend only on the additional correlation due to G3L, while the impact of GGL is removed by the compensated filter function.

As discussed in \citet{Schneider2003}, the imaginary part of the integral in equation (\ref{eq:NNMapFromG}), the B-mode $\expval{\mathcal{N}^2 M_\perp}$, is expected to vanish unless systematic effects cause a parity violation. We do not expect such a violation by any physical process; even the occurrence of B modes for the gravitational shear, which might be due to intrinsic alignments or clustering of source galaxies \citep{Schneider2002}, cannot induce a non-zero $\expval{\mathcal{N}^2 M_\perp}$. We nevertheless measure $\expval{\mathcal{N}^2 M_\perp}$, as a consistency check alongside the E mode $\expval{\mathcal{N}^2 M_\textrm{ap}}$ with $\tilde{\mathcal{G}}_\textrm{est}$. We only measure the aperture statistics for equal aperture scale radii $\theta$ and use the short-hand notations $\expval{\mathcal{N}^2 M_\textrm{ap}}(\theta, \theta, \theta) =: \expval{\mathcal{N}^2 M_\textrm{ap}}(\theta)$ and $\expval{\mathcal{N}^2 M_\perp}(\theta, \theta, \theta) =: \expval{\mathcal{N}^2 M_\perp}(\theta)$.


\section{Methods}
\label{sec:methods}

\subsection{Redshift weighting}
\label{sec:methods:redshift weighting}
To reduce the signal degradation by uncorrelated lens pairs, we define a redshift-weighted correlation function $\tilde{\mathcal{G}}_Z$, for which lens pairs are weighted according to their redshift difference $\delta z$. To this end, we introduce the redshift-weighting function $Z(\delta z),$ for which we choose a Gaussian,
\begin{equation}
  \label{eq:redshift_weighting_function}
  Z(\delta z) = \exp(-\frac{{\delta z}^2}{2\sigma_Z^2})\;.
\end{equation}
The width $\sigma_Z$ is a free parameter that should correspond to the typical redshift difference of correlated lens pairs. The weighting function is normalized such that it is unity if the galaxies have the same redshift. Averaging over the tangential ellipticities of lens-lens-source triplets weighted with $Z$ leads to an estimate of 
\begin{align}
\label{eq:estimatingSumGtildeZ}
&\frac{\int \dd{z_1} \int \dd{z_2}\, Z(\Delta z_{12})\, \expval{N(\va*{\vartheta}_1+\va*{\theta}, z_1)\, N(\va*{\vartheta}_2+\va*{\theta}, z_2)\, \gamma_\textrm{t}(\va*{\vartheta}_3+\va*{\theta})}}{\int \dd{z_1} \int \dd{z_2} Z(\Delta z_{12})\, \expval{N(\va*{\vartheta}_1+\va*{\theta}, z_1)\,N(\va*{\vartheta}_2+\va*{\theta}, z_2)}}\\
=:&\notag \frac{\tilde{\mathcal{G}}_Z(\vartheta_1, \vartheta_2, \phi)}{1+\omega_Z\left(|\va*{\vartheta}_1-\va*{\vartheta}_2|\right)}\;,
\end{align}
where $N(\va*{\vartheta}, z)$ is the number density of lens galaxies at angular position $\va*{\vartheta}$ and redshift $z$, and $\Delta z_{12} = z_1-z_2$. Equation~\eqref{eq:estimatingSumGtildeZ} defines the redshift-weighted correlation function $\tilde{\mathcal{G}}_Z$ and uses the redshift-weighted two-point angular correlation function $\omega_Z$. We estimate $\tilde{\mathcal{G}}_Z$ with
\begin{align}
  \label{eq:Gtilde_est_redsh  iftweighted}
  &\tilde{\mathcal{G}}_{Z, \text{est}}(b)\\
  &\notag=-\dfrac{\sum_{ijk} \, w_k \, \epsilon_k \, \textrm{e}^{-\textrm{i}(\varphi_{ik}+\varphi_{jk})} \,\left[1+\omega_Z\left(|\va*{\theta}_i - \va*{\theta}_j|\right)\right]\, Z(\Delta z_{ij}) \, \Delta_{ijk}(b)}{\sum_{ijk} w_k \,Z(\Delta z_{ij}) \,\Delta_{ijk}(b)}\;.
\end{align}
To estimate the redshift-weighted two-point correlation $\omega_Z$, we use the $N_\textrm{r}$ randoms, located at $\va*{\theta}_i'$, the $N_\textrm{d}$ lenses at the positions $\va*{\theta}_i$, and the estimator
\begin{equation}
\label{eq:LandySzalayModfied}
\omega_Z(\theta) = \frac{N_\textrm{r}^2 \, DD_Z(\theta)}{N_\textrm{d}^2 \, RR_Z(\theta)} - 2 \frac{N_\textrm{r} \, DR_Z(\theta)}{N_\textrm{d}\, RR_Z(\theta)} + 1\; ,
\end{equation}
with the modified pair-counts
\begin{align}
\label{eq:ModifiedPaircountsDDRR}
DD_Z(\theta) = \sum_{i=1}^{N_\textrm{d}} \sum_{j=1}^{N_\textrm{d}}&\Theta_\textrm{H}\left(\theta+{\Delta \theta}/{2} - |\va*{\theta}_i - \va*{\theta}_j|\right)\,\\
&\notag\times \Theta_\textrm{H}\left(-\theta+{\Delta \theta}/{2} + |\va*{\theta}_i - \va*{\theta}_j|\right)\, Z(\Delta z_{ij})\; , \\
RR_Z(\theta) = \sum_{i=1}^{N_\textrm{r}} \sum_{j=1}^{N_\textrm{r}}&\Theta_\textrm{H}\left(\theta+{\Delta \theta}/{2} - |\va*{\theta}_i - \va*{\theta}_j|\right)\\
&\notag\times \Theta_\textrm{H}\left(-\theta+{\Delta \theta}/{2} + |\va*{\theta}'_i - \va*{\theta}'_j|\right)\, Z(\Delta z_{ij})\; ,
\end{align}
and
\begin{align}
\label{eq:ModifiedPaircountsDR}
DR_Z(\theta) = \sum_{i=1}^{N_\textrm{d}} \sum_{j=1}^{N_\textrm{r}}& \Theta_\textrm{H}\left(\theta+{\Delta \theta}/{2} - |\va*{\theta}'_i - \va*{\theta}'_j|\right)\\
&\notag\times \Theta_\textrm{H}\left(-\theta+{\Delta \theta}/{2} + |\va*{\theta}'_i - \va*{\theta}'_j|\right)\, Z(\Delta z_{ij})\;.
\end{align}
Here, $\Theta_\textrm{H}$ is the Heaviside step function and $\Delta \theta$ is the bin size for which $\omega_Z$ is estimated. For $Z\equiv 1$, this estimator reduces to the standard Landy--Szalay estimator in Eq.~\eqref{eq:LandySzalay}.

The aperture statistics from the redshift-weighted correlation function $\tilde{\mathcal{G}}_Z$ are expected to have a higher S/N than the aperture statistics from the original $\tilde{\mathcal{G}}$. This expected improvement can be estimated with simplified assumptions. For this, we assume that the $N_\textrm{tot}$ lens-lens-source triplets can be split into $N_\textrm{true}$ physical triplets, each carrying the signal $s$, and $N_\textrm{tot}-N_\textrm{true}$ triplets carrying no signal. We further assume that all triplets carry the same uncorrelated noise $n$. Then, the measured total signal $S$, noise $N$ and S/N are
\begin{align}
\label{eq:SNR_unweighted}
S &= \frac{N_\textrm{true}}{N_\textrm{tot}} s, & N&=\frac{1}{\sqrt{N_\textrm{tot}}} n,  &\textrm{and }S/N&=\frac{N_\textrm{true}}{\sqrt{N_\textrm{tot}}}\frac{s}{n}\;.
\end{align}
With redshift weighting we decrease the effective number of triplets from $N_\textrm{tot}$ to $\tilde{N}_\textrm{tot}$, while retaining the same number of physical triplets $N_\textrm{true}$. The signal $\tilde{S}$, the noise $\tilde{N}$ and the new S/N $\tilde{S}/\tilde{N}$ are then
\begin{align}
\label{eq:SNR_weighted}
\tilde{S} &= \frac{N_\textrm{true}}{\tilde{N}_\textrm{tot}} s, & \tilde{N}&=\frac{1}{\sqrt{\tilde{N}_\textrm{tot}}} n,  &\textrm{and }\tilde{S}/\tilde{N}&=\frac{N_\textrm{true}}{\sqrt{\tilde{N}_\textrm{tot}}}\frac{s}{n}\;.
\end{align}

Consequently, redshift weighting increases the noise by a factor of ${({N_\textrm{tot}}/{\tilde{N}_\textrm{tot}})}^{1/2}$. Nonetheless, the S/N improves by ${({N_\textrm{tot}}/{\tilde{N}_\textrm{tot}})}^{1/2}$ because the signal increases by ${N_\textrm{tot}}/{\tilde{N}_\textrm{tot}}$. Accordingly, we expect the S/N to increase approximately by the square root of the signal increase.

The critical parameter for the redshift weighting is the width $\sigma_z$ of the weighting function. For our application on the observational and simulated data described in Sect.~\ref{sec:data}, we choose $\sigma_z = 0.01$. Because lens pairs that carry signal and those that do not are not clearly divided, the choice of this parameter needs to remain somewhat arbitrary. However, three arguments can be made to motivate our choice.

The first argument considers the galaxy correlation length. \citet{Farrow2015} measured the two-point correlation function of galaxies in the Galaxy and Mass Assembly survey (GAMA) and found correlation lengths between $3.28 \pm 0.42 \, h^{-1}\, \textrm{Mpc}$ and $38.17 \pm 0.47 \, h^{-1}\, \textrm{Mpc}$, depending on the stellar masses of the galaxies. The same function was measured by \citet{Zehavi2011} in the Sloan Digital Sky Survey (SDSS). They found similar correlation lengths between $4.2\,h^{-1}\, \textrm{Mpc}$ and $10.5\,h^{-1}\, \textrm{Mpc}$. These correlation lengths correspond to redshift differences between $0.001$ and $0.005$ at the median redshift of GAMA of $z=0.21$ . We assume that galaxies separated by more than twice the correlation length are only weakly correlated, and therefore our choice of $\sigma_z=0.01$ seems appropriate.

\begin{figure}
\resizebox{\hsize}{!}{\includegraphics[width=\linewidth]{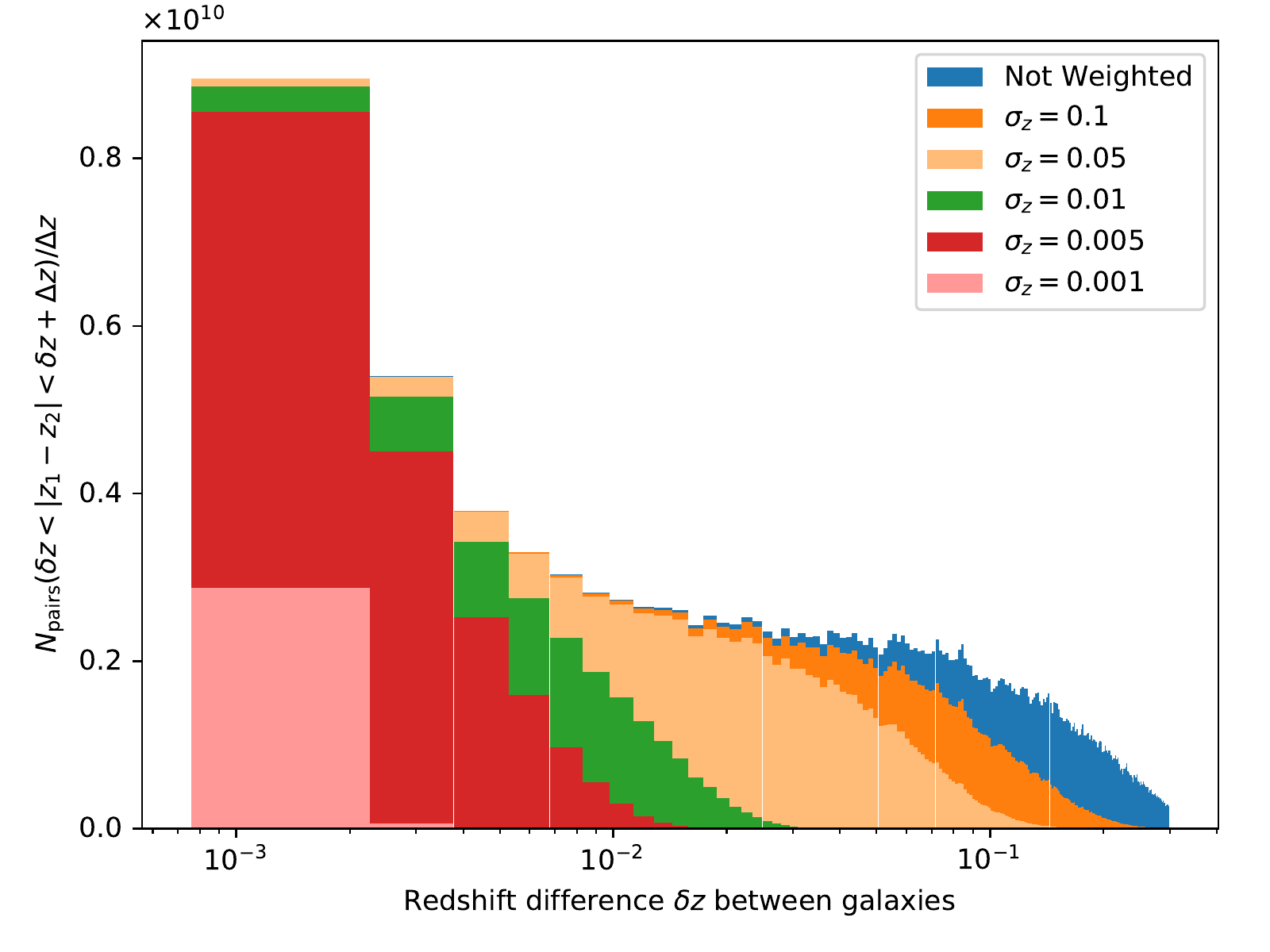}}
\caption{Weighted number of lens galaxy pairs in our sample from the MR with fixed angular separation between $\ang[astroang]{;4.5;}$ and $\ang[astroang]{;5.5;}$ per redshift difference between the pairs. Different colours indicate different widths of the Gaussian weighting function. The blue histogram shows the unweighed distribution, and the green histogram shows the distribution with the weighting chosen for the G3L measurements.}
\label{fig:redshiftdifference_distribution}
\end{figure}

The second argument relates to the distribution of lens galaxy pairs with their redshift difference. The blue histogram in Fig.~\ref{fig:redshiftdifference_distribution} shows the number of galaxy pairs per redshift difference $\delta z$ with fixed angular separation between $\ang[astroang]{;4.5;}$ and $\ang[astroang]{;5.5;}$ in our lens sample from the MR (see Sect.~\ref{sec:data}). This distribution has a prominent peak for small $\delta z$ and a broad background distribution. Thus, most galaxy pairs that appear close on the sky are also close in redshift space. These physical pairs make up the peak. However, the background distribution shows that there are also many galaxy pairs with small angular separation whose redshift difference is large. The redshift weighting function should now be chosen in such a way that pairs inside the peak are preserved, while the background is suppressed.

The other histograms in Fig.~\ref{fig:redshiftdifference_distribution} show different weighted distributions, where the number of galaxy pairs is multiplied by the redshift-weighting function from Eq.~\eqref{eq:redshift_weighting_function}. This gives the effective number of galaxy pairs per redshift difference bin that are considered for the improved $\tilde{\mathcal{G}}$ estimator. Here, the effect of different $\sigma_z$ is visible. The peak is preserved when we use $\sigma_z=0.1$ and $0.05$ , but a high percentage of the background is still present in the weighted distribution. For $\sigma_z=0.005$ and $\sigma_z=0.001$, the background is removed, but parts of the peak are also suppressed. A middle ground is found for $\sigma_z=0.01$. Here, the tails of the peak still contribute, whereas most of the background galaxy pairs are suppressed. Consequently, we adopt this value for the measurement of $\tilde{\mathcal{G}}$ and subsequently $\expval{\mathcal{N}^2 M_\textrm{ap}}$.

The third argument for our choice of $\sigma_z$ considers the peculiar velocities of galaxies in clusters, which can cause redshift differences of correlated galaxy pairs inside the same halo. The weighting function $Z$ needs to be broad enough to avoid discarding galaxy pairs whose redshift differences are induced simply by their peculiar motion. Velocities of galaxies inside halos can reach up to $\SI{1000}{\kilo\metre\per\second}$, leading to redshift differences of up to $0.006$. This is a lower bound for $\sigma_z$, therefore choosing $\sigma_z=0.01$ appears valid.

\subsection{New binning scheme}
\label{sec:methods:binning scheme}
In previous work \citep{Simon2008, Simon2013}, $\tilde{\mathcal{G}}$ was measured on a regular grid with logarithmic spacing in the lens-source separations $\vartheta_1$ and $\vartheta_2$ and linear spacing in the opening angle $\phi$. The aperture statistics were then calculated by summing over this grid.

However, in this approach, the estimator for $\tilde{\mathcal{G}}$ is undefined in any bin for which no triplet was found. In previous work, $\tilde{\mathcal{G}}$ was therefore set to zero in these empty bins. As a result $\expval{\mathcal{N}^2 M_\textrm{ap}}$, which is obtained by integrating over the estimated $\tilde{\mathcal{G}}$, was underestimated \citep{Simon2008}. This bias occurs for both small and large scales: At small scales, the bins for $\vartheta_1$ and $\vartheta_2$ are small because of the logarithmic binning, therefore many bins remain empty. At large scales, certain bins automatically remain empty because the opening angle $\phi$ cannot assume all values between $0$ and $2\pi$ if $\vartheta_1$ or $\vartheta_2$ are larger than the side length of the field of view.

The bin sizes and number of lens-lens-source triplets affect by how much $\expval{\mathcal{N}^2 M_\textrm{ap}}$ is underestimated. If the bins are smaller, the probability of encountering empty bins is higher and the bias is stronger. If the number of triplets increases, there are fewer empty bins and the bias decreases.

\begin{figure}
\centering
\resizebox{\hsize}{!}{\includegraphics{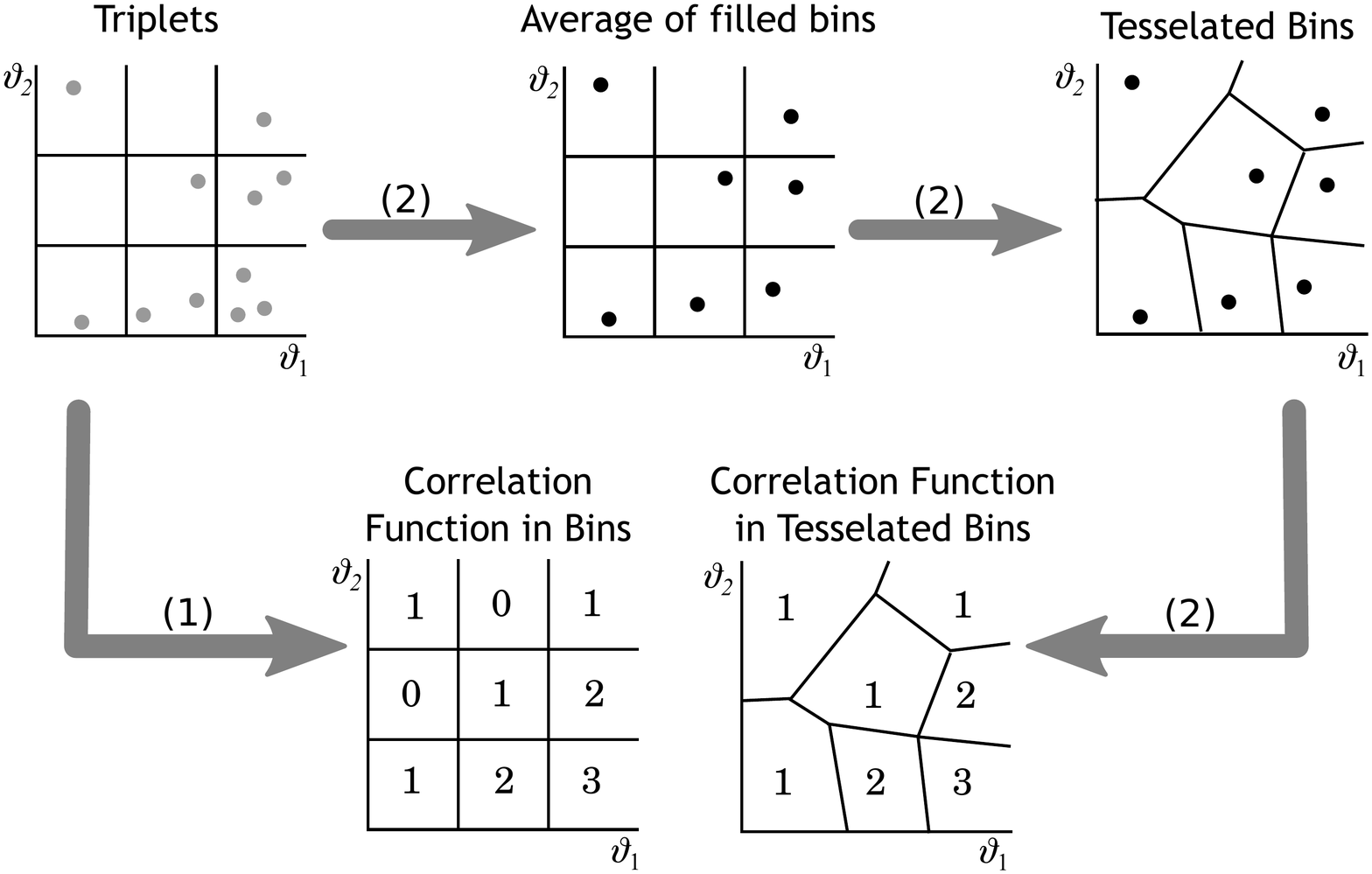}}
\caption{Illustration of the old (1) and new (2) binning scheme for the calculation of $\tilde{\mathcal{G}}$. In the old binning scheme, $\tilde{\mathcal{G}}$ was calculated directly from the lens-lens-source triplets inside a given bin. In the new binning scheme, the average of the lens-lens-source triplets in a bin is calculated first. These averages are used as seeds for a Voronoi tessellation of the parameter space. Each Voronoi cell is then considered as a new bin for which $\tilde{\mathcal{G}}$ is estimated. The aperture statistics are obtained by integrating over the new bins. We show only two dimensions here, but for the measurement the tessellation was also made along the third parameter $\phi$.}
\label{fig:Tesselation}
\end{figure}

To account for this effect, we introduce an adaptive binning scheme, illustrated in Fig.~\ref{fig:Tesselation}. In this new scheme, bins are defined such that they contain at least one triplet, and therefore the estimator for $\tilde{\mathcal{G}}$ is always well defined. For this, $\tilde{\mathcal{G}}$ is first estimated on a regular grid, together with the average side lengths of the triplets in each bin. Then, in all bins for which triplets are found, the measured $\tilde{\mathcal{G}}$ is associated with the average $\vartheta_1$, $\vartheta_2$ and $\phi$ of the corresponding bin. We use the averages of the triplets in filled bins as seeds to divide the parameter space by a Voronoi tessellation, using the library \texttt{voro++} by \citet{Voro++}. Each Voronoi cell is then considered as a new bin for which $\tilde{\mathcal{G}}$ is estimated. These bins
by definition contain at least one triplet. The aperture statistics are obtained by integrating over the $N_\textrm{bin}$ new bins, using the numerical approximation of Eq.~\eqref{eq:NNMapFromG},
\begin{align}
&\expval{\mathcal{N}^2 M_\textrm{ap}}(\theta) + \textrm{i}\, \expval{\mathcal{N}^2 M_\perp}(\theta) = \sum_{i=1}^{N_\textrm{bin}} V(b_i)\, \tilde{\mathcal{G}}_\textrm{est}(b_i) \, A_{NNM}(b_i|\theta)\; ,
\end{align}
where $b_i$ is the $i$th bin, $V(b_i)$ is the volume of this bin, and $A_{NNM}(b_i|\theta)$ is the kernel function of Eq.~\eqref{eq:NNMapFromG} evaluated at the seed of $b_i$. We estimate $\tilde{\mathcal{G}}$ on a grid with $128 \times 128 \times 128$ bins with $\vartheta_1$ and $\vartheta_2$ between $\ang[astroang]{;0.15;}$ and $\ang[astroang]{;320;}$ for the data based on the MR (see Sect.~\ref{sec:data:MR}) and between $\ang[astroang]{;0.15;}$ and $\ang{;200;}$ for the simple mock data (see Sect.~\ref{sec:data:simpleMockData}). The tessellation reduces the number of bins by approximately \SI{3}{\percent} in both cases.

\subsection{Conversion into physical units}
\label{sec:methods:physical units}
With the lens redshifts $z_1$ and $z_2$, we can transform the projected angular separation vectors $\va*{\vartheta}_1$ and $\va*{\vartheta}_2$ into physical separations $\va{r}_1$ and $\va{r}_2$ on a plane midway between the two lenses, using
\begin{equation}
\va{r}_{1,2} = D_\textrm{A}\left(0,z_{12}\right)\, \va*{\vartheta}_{1,2}=: D_\textrm{A}(z_{12})\,\va*{\vartheta}_{1,2} \; ,
\end{equation}
with the angular diameter distance $D_\textrm{A}(z_a,z_b)$ between redshifts $z_a$ and $z_b$ and the average lens redshift $z_{12}=(z_1+z_2)/2$.

The correlation function $\tilde{\mathcal{G}}_Z$ can therefore be estimated in physical scales in the bin $b$ of ${r}_1$, ${r}_2$ and $\phi$ as
\begin{align}
\label{eq:Gtilde_est_phys_noSigCrit}
&\tilde{\mathcal{G}}_{Z, \textrm{est}}(b) \\
=&\notag -\frac{\sum_{ijk} w_k \, \epsilon_k \, \textrm{e}^{-\textrm{i} (\varphi_{ik} + \varphi_{jk})}\, \left[1 + \omega\left(|\va*{\theta}_i - \va*{\theta}_j|\right)\right] Z(\Delta z_{ij}) \Delta_{ijk}^{\textrm{ph}}(b)}{\sum_{ijk} w_k \, Z(\Delta z_{ij})\, \Delta_{ijk}^{\textrm{ph}}(b)}\; ,
\end{align}
with
\begin{align}
\label{eq:trianglePhys}
&\Delta_{ijk}^{\textrm{ph}}(b) = 
\begin{cases}
1 &\textrm{for }\left(D_A(z_{ij})\,|\va*{\theta}_k-\va*{\theta}_i|,D_A(z_{ij})\,|\va*{\theta}_k-\va*{\theta}_j|,\phi_{ijk}\right) \in b \\
0 &\textrm{otherwise.}
\end{cases}
\end{align}
This $\tilde{\mathcal{G}}_Z$ still depends on the redshift distribution of sources because the gravitational shear $\gamma_\textrm{t}$ depends on the lensing efficiency, which in turn depends on the distances between observer and source and lens and source. To compare the measurements of different surveys with varying source redshift distributions, it is therefore useful to correlate the galaxy number density not with the tangential shear $\gamma_\textrm{t}$, but instead with the projected excess mass density $\Delta \Sigma$, given by
\begin{equation}
\label{eq:excessmassdensity}
\Delta \Sigma(\va*{\theta}, z_\textrm{d}, z_\textrm{s}) =
\begin{cases}
 \dfrac{\gamma_{\rm t}(\va*{\theta})}{\Sigma_{\textrm{crit}}^{-1}(z_\textrm{d}, z_\textrm{s})}\;\;&\textrm{for}\;z_\textrm{d}<z_\textrm{s} \\
 0 \;\; &\textrm{else}
\end{cases}\; ,
\end{equation}
with the inverse critical surface mass density
\begin{equation}
  \label{eq:Sigma_crit}
  \Sigma_{\textrm{crit}}^{-1}(z_\textrm{d}, z_\textrm{s}) = \dfrac{4\,\pi\, G}{c^2} \dfrac{D_A(z_\textrm{d}, z_\textrm{s})\, D_A(z_\textrm{d})} {D_A(z_\textrm{s})}\,\Theta_\textrm{H}(z_\textrm{s} - z_\textrm{d})\;.
\end{equation}
Thus, we are interested in the correlation function $\tilde{\mathcal{G}}_\textrm{phys}$, defined by
\begin{align}
\label{eq:definitionGtildePhys}
&\notag \tilde{\mathcal{G}}_\textrm{phys}(\va{r_1}, \va{r_2})\\
 &=\frac{1}{\overline{N}^2}\int \dd{z_1}\, \int \dd{z_2}\, p(z_1)\, p(z_2)\, Z(\Delta z_{12})\\
&\notag \quad\times \expval{N\left(\frac{\va{r}_1}{D_\textrm{A}(z_{12})} + \va*{\theta}, z_1\right)\, N\left(\frac{\va{r}_2}{D_\textrm{A}(z_{12})} + \va*{\theta}, z_2 \right)\, \Delta \Sigma(\va*{\theta})}\\
\notag &=: \tilde{\mathcal{G}}_\textrm{phys}(r_1, r_2, \phi)\;.
\end{align}
To estimate this quantity with a maximum likelihood estimator, we need to multiply the weight $w_k$ of each source galaxy with $\Sigma_{\textrm{crit}}^{-2}$ \citep{Sheldon2004}. This leads to the estimator
\begin{align}
\label{eq:GtildePhys_precise}
&\tilde{\mathcal{G}}_\textrm{est,phys}(b)=\\
&-\notag\dfrac{\sum\limits_{ijk} w_k\, \epsilon_k \, \textrm{e}^{-\textrm{i}(\varphi_{ik}+\varphi_{jk})}\left[1 + \omega\left(|\va*{\theta}_i - \va*{\theta}_j|\right)\right]Z(\Delta z_{ij})\,{\Sigma_{\textrm{crit}}^{-1}}_{ijk}\,\Delta_{ijk}^{\textrm{ph}}(b)}{\sum\limits_{ijk} w_k \, \Sigma_{\textrm{crit}}^{-2}(z_\textrm{d}, z_\textrm{s})\,  Z(\Delta z_{ij})\, \Delta_{ijk}^{\textrm{ph}}(b)}\;,
\end{align}
with ${\Sigma_{\textrm{crit}}}_{ijk}=\Sigma_{\textrm{crit}}(z_\textrm{ij}, z_\textrm{k})$. 

This estimator requires a precise knowledge of the source redshifts. For the application to real data, however, only photometric redshift estimates are often available for source galaxies. Therefore, we do not use the exact $\Sigma_{\textrm{crit}}^{-1}$ for each triplet, but instead $\bar{\Sigma}_{\textrm{crit}}^{-1}$, which is averaged over the source distribution $p_\textrm{s}(z_\textrm{s})$ as
\begin{equation}
  \label{eq:Sigma_crit_ave}
  \bar{\Sigma}_{\textrm{crit}}^{-1}(z_\textrm{d})=\int \dd{z_\textrm{s}}\, p_\textrm{s}(z_\textrm{s})\, \Sigma_{\textrm{crit}}^{-1}(z_\textrm{d}, z_\textrm{s})\;.
\end{equation}
Consequently, we estimate $\tilde{\mathcal{G}}_\textrm{phys}$ with
\begin{align}
\label{eq:3ptcorrelation_estimator_physical}
&\tilde{\mathcal{G}}_\textrm{est,phys}(b)=\\
&\notag - \dfrac{\sum\limits_{ijk} w_k \, \epsilon_k \, \textrm{e}^{-\textrm{i}(\varphi_{ik}+\varphi_{jk})}\left[1 + \omega\left(|\va*{\theta}_i - \va*{\theta}_j|\right)\right]  Z(\Delta z_{ij})\, \bar{\Sigma}_{\textrm{crit}}^{-1}(z_{ij})\, \Delta_{ijk}^{\textrm{ph}}(b)}{\sum\limits_{ijk} w_k \, \bar{\Sigma}_{\textrm{crit}}^{-2}(z_{ij})\,  Z(\Delta z_{ij})\, \Delta_{ijk}^{\textrm{ph}}(b)}\;.
\end{align}
We convert this physical three-point correlation function into physical aperture statistics with
\begin{align}
\label{eq:NNMapPhysDefinition}
&\notag\expval{\mathcal{N}^2 M_\textrm{ap}}_\textrm{phys} (R) + \textrm{i}\, \expval{\mathcal{N}^2 M_\perp}_\textrm{phys} (R) \\
&=  \int_0^{\infty}\dd{r_1}\, r_1\, \int_0^{\infty}\dd{r_2}\, r_2\, \int_0^{2\pi} \dd{\phi}\,  \tilde{\mathcal{G}}_\textrm{phys}(r_1, r_2, \phi)\\
&\notag \quad \times {A}_{\mathcal{N}\mathcal{N}M}\left[D_A^{-1}(z_{12})\,r_1, D_A^{-1}(z_{12})\,r_2,\phi\; |\; D_A^{-1}(z_{12})\,R\right]\;.
\end{align}
These aperture statistics are in units of mass over area.

\subsection{Magnification of lens galaxies}
\label{sec:methods:magnification bias}
Magnification of lens galaxies by the LSS affects G3L because the apparent magnitude and number density of lenses is changed \citep{Bartelmann2001}. In the weak-lensing limit, the number density of lens galaxies at angular position $\va*{\vartheta}$ and redshift $z$ is changed from the intrinsic number density $N_0(\va*{\vartheta}, z)$ to
 \begin{equation}
 N(\va*{\vartheta}, z) = N_0(\va*{\vartheta}, z) + 2\,[\alpha(z) - 1]\, \overline{N} \, \kappa(\va*{\vartheta}, z)\;,
 \end{equation}
 where $\kappa(\va*{\vartheta}, z)$ is the convergence caused by all matter in front of redshift $z$, and $\alpha(z)$ is the negative slope of the luminosity function $\Phi(S, z)$ at the flux limit $S_\textrm{lim}$ of lens galaxies. We define $\alpha$ by 
 \begin{equation}
\alpha = -\frac{\dd{\ln \Phi}}{\dd{\ln S}}\left(S_\textrm{lim}\right)\;.
 \end{equation}
Consequently, the correlation function $\tilde{\mathcal{G}}_Z$ with the effect of lens magnification is
\begin{align}
&\tilde{\mathcal{G}}_Z(\vartheta_1, \vartheta_2, \phi)=\\
&\notag \int \dd{z_1}\, \int \dd{z_2}\, Z(z_1-z_2)\\
 &\notag \times \Biggl\lbrace \frac{1}{\overline{N}^2}\,\expval{N_0(\va*{\vartheta}_1+\va*{\theta}, z_1)\,N_0(\va*{\vartheta}_2+\va*{\theta}, z_2)\,\gamma_\textrm{t}(\va*{\theta})}\\
&\notag \quad \left. + \frac{2 [\alpha(z_2) - 1]}{\overline{N}}\, \expval{N_0(\va*{\vartheta}_1+\va*{\theta}, z_1)\,\kappa(\va*{\vartheta}_2+\va*{\theta}, z_2)\,\gamma_\textrm{t}(\va*{\theta})}\right.\\
&\notag \quad \left. + \frac{2[\alpha(z_1) - 1]}{\overline{N}}\, \expval{\kappa(\va*{\vartheta}_1+\va*{\theta}, z_1)\,N_0(\va*{\vartheta}_2+\va*{\theta}, z_2)\,\gamma_\textrm{t}(\va*{\theta})}\right.\\
&\notag \quad  + \frac{4[\alpha(z_1) - 1][\alpha(z_2)-1]}{\overline{N}}\, \expval{\kappa(\va*{\vartheta}_1+\va*{\theta}, z_1)\,\kappa(\va*{\vartheta}_2+\va*{\theta}, z_2)\,\gamma_\textrm{t}(\va*{\theta})}\Biggr\rbrace\;.
\end{align}
With the intrinsic aperture number count 
\begin{align}
&\mathcal{N}_{0,\theta}(\vec{\vartheta}, z) = \frac{1}{\overline{N}}\,\int \dd[2]{\vartheta'}\, U_\theta(|\vec{\vartheta} - \vec{\vartheta}'|)\, N_{0}\left(\vec{\vartheta}', z\right)\,,
\end{align}
and $M_{\textrm{ap}, \theta}$ as defined in Eq.~\eqref{eq:DefinitionApertureMass}, the aperture statistics are
\begin{align}
\label{eq:NNMapWithMagBias}
&\expval{\mathcal{N}^2 M_\textrm{ap}}(\theta)=\\
&\notag \int \dd{z_1}\, \int \dd{z_2}\, Z(z_1-z_2)\\
&\notag \times \Bigl\lbrace \expval{\mathcal{N}_{0,\theta}(\vec{\vartheta}, z_1)\, \mathcal{N}_{0,\theta}(\vec{\vartheta}, z_2)\, M_{\textrm{ap},\theta}(\vec{\vartheta})}\\
&\notag \quad \left. + 2[\alpha(z_2)-1]\, \expval{\mathcal{N}_{0,\theta}(\vec{\vartheta}, z_1)\, M_{\textrm{ap}, \theta}(\vec{\vartheta}, z_2) \, M_{\textrm{ap},\theta}(\vec{\vartheta})}\right.\\
&\notag \quad \left. + 2[\alpha(z_1)-1]\, \expval{M_{\textrm{ap}, \theta}(\vec{\vartheta}, z_1) \,\mathcal{N}_{0,\theta}(\vec{\vartheta}, z_2)\,  M_{\textrm{ap},\theta}(\vec{\vartheta})}\right. \\
&\notag  \quad + 4[\alpha(z_1)-1][\alpha(z_2)-1]\expval{M_{\textrm{ap}, \theta}(\vec{\vartheta}, z_1)\,M_{\textrm{ap}, \theta}(\vec{\vartheta}, z_2)\, M_{\textrm{ap}, \theta}(\vec{\vartheta})   }\Bigr\rbrace.
\end{align}
Thus, the measured aperture statistics do not only include the intrinsic first term, but three additional terms that are due to lens magnification. These lens magnification terms, however, can be measured using as redshift-weighting function $Z$ not a Gaussian, but a step function,
\begin{equation}
\label{eq:stepfunction weighting}
Z(z_1-z_2) = \Theta_\textrm{H}(z_2-z_1-\Delta z).
\end{equation}
This means that only lens pairs with a redshift difference larger than $\Delta z$ and $z_2>z_1$ are counted in the estimator in Eq.~\eqref{eq:Gtilde_est_redsh  iftweighted}. As explained in Sect.~\ref{sec:methods:redshift weighting}, we expect lens pairs with redshift differences larger than $0.01$ to be intrinsically uncorrelated. When we choose $\Delta z =0.01$, the first term in Eq.~\eqref{eq:NNMapWithMagBias}, which contains only the correlation of intrinsic number densities, should vanish. The measured $\expval{\mathcal{N}^2 M_\textrm{ap}}$ is then purely the correlation due to the lens magnification. We measure this $\expval{\mathcal{N}^2 M_\textrm{ap}}$ with the estimator in Eq.~\eqref{eq:Gtilde_est_redsh  iftweighted}, using the step function weighting. If this signal is then subtracted from the measured $\expval{\mathcal{N}^2 M_\textrm{ap}}$ of all lenses, we obtain the intrinsic aperture statistics.

Because we tested our approach on simulated data from the MR, for which both the number density and convergence are available at different redshift planes, we can also use another approach to measure the terms due to lens magnification. In this approach we use the relation of the observed aperture number count $\mathcal{N}_\theta$ to the intrinsic aperture number count $\mathcal{N}_{0, \theta}$ and aperture mass $M_{\textrm{ap}, \theta}$ through
\begin{align}
\mathcal{N}_{\theta}(\vec{\vartheta}, z) &= \frac{1}{\overline{N}}\,\int \dd[2]{\vartheta'}\, U_\theta(\vec{\vartheta} - \vec{\vartheta}'|)\, N\left(\vec{\vartheta}', z\right)\\
& = \mathcal{N}_{0, \theta} + 2\,[\alpha(z)-1]\, M_{\textrm{ap}, \theta}(\va*{\vartheta}, z)\;.
\end{align}

Consequently, Eq.~\eqref{eq:NNMapWithMagBias} with the step function weighting in Eq.~\eqref{eq:stepfunction weighting} leads to
\begin{align}
\label{eq:NNMapWithMagBias2}
&\expval{\mathcal{N}^2 M_\textrm{ap}}(\theta)\\
&=\notag\int_0^{z_{\rm max}}\, \dd{z_1} \int_{z_1+\Delta z}^{z_{\rm max}} \dd{z_2}\, \left\lbrace \expval{\mathcal{N}_{0,\theta}(\vec{\vartheta}, z_1)\, \mathcal{N}_{0,\theta}(\vec{\vartheta}, z_2)\, M_{\textrm{ap},\theta}(\vec{\vartheta})}\right.\\
&\quad \notag \left. + 2\,[\alpha(z_2)-1]\, \expval{\mathcal{N}_{\theta}(\vec{\vartheta}, z_1)\, M_{\textrm{ap}, \theta}(\vec{\vartheta}, z_2) \, M_{\textrm{ap},\theta}(\vec{\vartheta})}\right.\\
&\quad\notag \left. + 2\,[\alpha(z_1)-1]\, \expval{M_{\textrm{ap}, \theta}(\vec{\vartheta}, z_1) \,\mathcal{N}_{\theta}(\vec{\vartheta}, z_2)\,  M_{\textrm{ap},\theta}(\vec{\vartheta})}\right. \\
&\quad\notag \left. - 4\,[\alpha(z_1)-1]\,[\alpha(z_2)-1] \expval{M_{\textrm{ap}, \theta}(\vec{\vartheta}, z_1)\,M_{\textrm{ap}, \theta}(\vec{\vartheta}, z_2)\, M_{\textrm{ap}, \theta}(\vec{\vartheta})   }\right\rbrace,
\end{align}
where the terms due to lens magnification are given by the observed instead of by the intrinsic aperture number count. For a numerical evaluation, the integrals can be converted into sums over $M$ redshift slices, so
\begin{align}
&\notag \expval{\mathcal{N}^2 M_\textrm{ap}}(\theta)\\
&=\sum_{i=0}^{M}\sum_{j=i+1}^M \Delta z_i \, \Delta z_j \left\lbrace \expval{\mathcal{N}_{0,\theta}(\vec{\vartheta}, z_i)\, \mathcal{N}_{0,\theta}(\vec{\vartheta}, z_j)\, M_{\textrm{ap},\theta}(\vec{\vartheta})}\right.\\
&\notag \quad \left. + 2\,[\alpha(z_j)-1]\, \expval{\mathcal{N}_{\theta}(\vec{\vartheta}, z_i)\, M_{\textrm{ap}, \theta}(\vec{\vartheta}, z_j) \, M_{\textrm{ap},\theta}(\vec{\vartheta})}\right.\\
&\notag  \quad\left. + 2\,[\alpha(z_i)-1] \, \expval{M_{\textrm{ap}, \theta}(\vec{\vartheta}, z_i) \,\mathcal{N}_{\theta}(\vec{\vartheta}, z_j)\,  M_{\textrm{ap},\theta}(\vec{\vartheta})}\right. \\
&\notag \quad\left. - 4\,[\alpha(z_i)-1][\alpha(z_j)-1]\, \expval{M_{\textrm{ap}, \theta}(\vec{\vartheta}, z_i)M_{\textrm{ap}, \theta}(\vec{\vartheta}, z_j)\, M_{\textrm{ap}, \theta}(\vec{\vartheta})   }\right\rbrace\\
\label{eq:NNMapWithMagBiasNum}
&=:\sum_{i=0}^{M}\sum_{j=i+1}^M \Delta z_i \, \Delta z_j \, \expval{\mathcal{N}_{0,\theta}(\vec{\vartheta}, z_i)\, \mathcal{N}_{0,\theta}(\vec{\vartheta}, z_j)\,M_{\textrm{ap},\theta}(\vec{\vartheta})}\\
&\notag \quad + L_{\mathcal{NMM}}(\theta) + L_{\mathcal{MNM}}(\theta)+L_\mathcal{MMM}(\theta)\; .
\end{align}

Using Eq.~\eqref{eq:NNMapWithMagBiasNum}, we measure the lens magnification terms $L_\mathcal{MMM}$, $L_\mathcal{MNM}$ and $L_\mathcal{NMM}$ directly in the simulated data based on the MR for $z_j<0.5$ and $z_i<z_j$. 

For this, we first convolve the number density and convergence maps at each redshift plane with the filter function $U_\theta$ to obtain $\mathcal{N}_{\theta}(\vec{\vartheta}, z_i)$ and $M_{\textrm{ap}, \theta}(\vec{\vartheta}, z_j)$. We then multiply the aperture statistics for each combination of $z_i$ and $z_j$ and spatially average the products to obtain
$\expval{\mathcal{N}_{\theta}(\vec{\vartheta}, z_i)\, M_{\textrm{ap}, \theta}(\vec{\vartheta}, z_j)\,  M_{\textrm{ap},\theta}(\vec{\vartheta})}$ and $\expval{M_{\textrm{ap}, \theta}(\vec{\vartheta}, z_i)\,M_{\textrm{ap}, \theta}(\vec{\vartheta}, z_j)\, M_{\textrm{ap}, \theta}(\vec{\vartheta})   }$. These averages are then multiplied by the appropriate $\alpha$ and summed over. We repeat this procedure for different aperture scale radii $\theta$ between $\ang[astroang]{;0.5;}$ and $\ang[astroang]{;8;}$.

For this calculation, the slope $\alpha(z)$ of the lens luminosity function needs to be known. To obtain $\alpha(z)$, we extract the luminosity function $\Phi(S,z)$ at each redshift plane of the MR, with $S$ measured in the $r$-band filter. We then fit a power law to $\Phi(S,z)$ in the proximity of the limiting flux. This flux is given in our case by the limiting $r$-band magnitude, chosen to be $r_\textrm{lim}=19.8\,\textrm{mag}$. The slopes for each redshift $z$ are the $\alpha(z)$ given in Table~\ref{tab:alpha}.

\begin{table}
        \caption{Slopes $\alpha(z)$ of the luminosity function at different redshifts $z$ in the MR. The limiting magnitude of galaxies is $r_\textrm{lim}=19.8\,\textrm{mag}$.}
        \label{tab:alpha}
        \centering
\begin{tabular}{ll}
        \hline
        $z$ & $\alpha(z)$\\ \hline
        0.46 & 2.51 \\
        0.41 & 2.38 \\
        0.36 & 2.01 \\
        0.32 & 1.80 \\
        0.28 & 1.36 \\
        0.24 & 1.15 \\
        0.21 & 0.91 \\
        0.17 & 0.78 \\
        0.14 & 0.49 \\
        0.12 & 0.48 \\
        0.09 & 0.48 \\
        0.06 & 0.47 \\
        0.04 & 0.17\\
        \hline
\end{tabular}   
\end{table}
\section{Data}
\label{sec:data}

\subsection{Simulated data based on the MR}
\label{sec:data:MR}
We tested our new estimator with simulated data sets from the MR. The MR \citep{Springel2005} is a dark matter-only cosmological N-body-simulation. It traces the evolution of $2160^3$ dark matter particles of mass $m=8.6\times 10^8\,h^{-1}\,\textrm{M}_\odot$  from redshift $z=127$ to today in a cubic region with co-moving side length $500\, h^{-1}\,\textrm{Mpc}$. For this, a flat $\Lambda$CDM cosmology was assumed, with matter density $\Omega_\textrm{m}=0.25$, baryon density $\Omega_\textrm{b}= 0.045$, dark energy density $\Omega_\Lambda= 0.75$, Hubble constant $H_0= 73 \,\textrm{km}\,\textrm{s}^{-1}\,\textrm{Mpc}^{-1}$ , and power spectrum normalization $\sigma_8=0.9$.

\begin{figure}
        \resizebox{\hsize}{!}{\includegraphics[width=\linewidth]{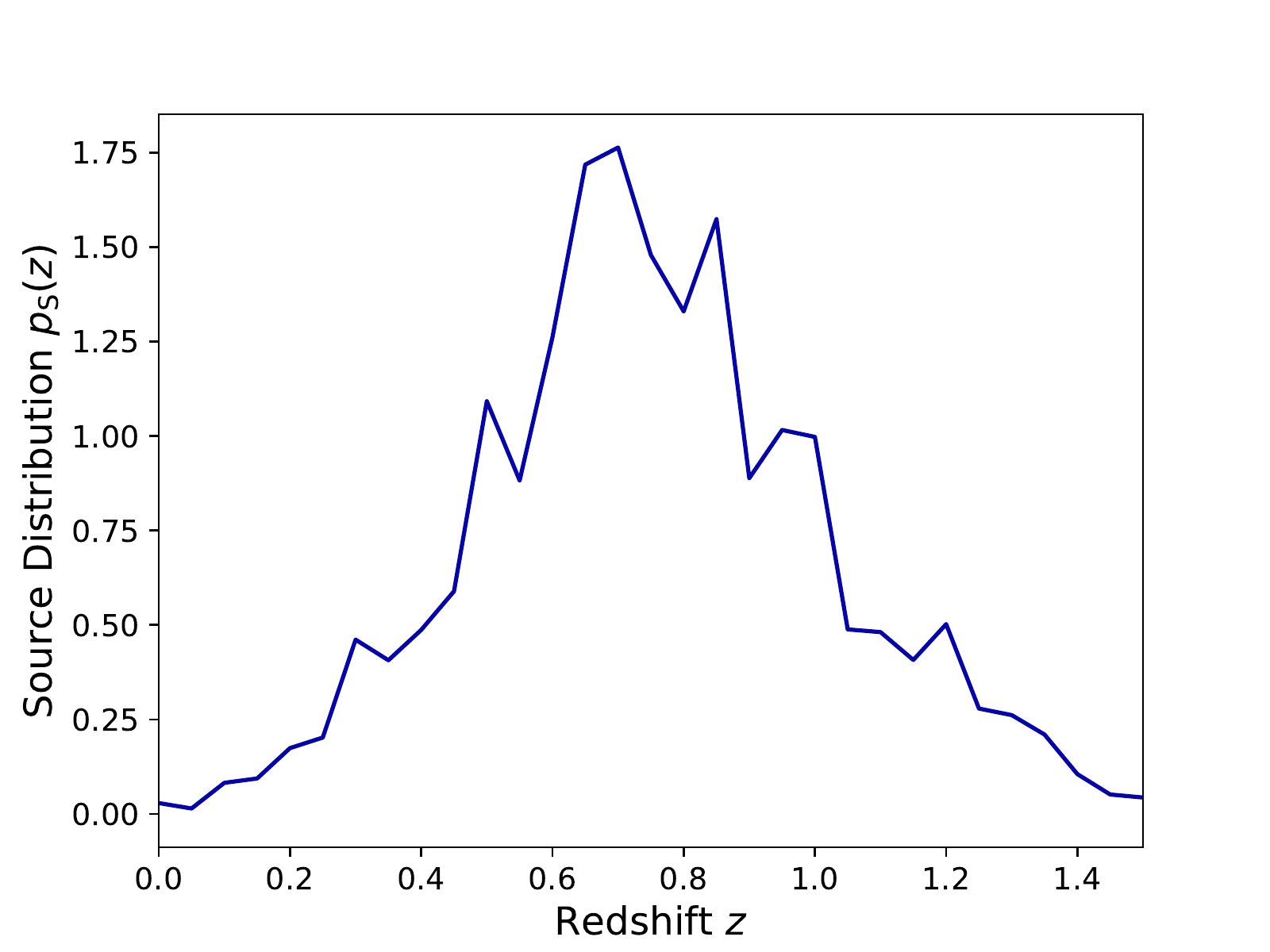}}
        \caption{Assumed source redshift distribution. This distribution is used to weigh the shear maps at different redshifts in the MR. It is modelled after the redshift distribution of galaxies in \citet{Hildebrandt2018}.}
        \label{fig:n_z}
\end{figure}

Using the multiple-lens-plane ray-tracing algorithm by \citet{Hilbert2009}, we created maps of the complex gravitational shear $\gamma$ caused by the matter distribution for a set of source redshift planes. For each redshift, 64 maps of $\gamma$ on a regular mesh with $4096^2$ pixels, corresponding to $4 \times 4\, \textrm{deg}^2$, were obtained. We combined the shear of nine different redshifts between $z=0.5082$ and $z=1.1734$ by summing $\gamma$ weighted by an assumed source redshift distribution $p_\textrm{s}(z)$. This redshift distribution, shown in Fig.~\ref{fig:n_z}, was modelled after the redshift distribution of galaxies in the Kilo-Degree Survey (KiDS; \citealp{Wright2018, Hildebrandt2018}). To mimic the shape noise in observational data, we added a random number drawn from a Gaussian probability distribution with standard deviation 0.3 to both shear components at each pixel. This gave us 64 maps of mock source galaxies.

Lens galaxies in the simulation were created by using the SAM by \citetalias{Henriques2015}. It is one of various SAMs that have been implemented on the MR \citep[see e.g.][]{Guo2011, Bower2006}, but as \citet{Saghiha2017} have shown, this model agrees particularly well with measurements of GGL and G3L in CFHTLenS. To simulate the selection function of observations, we applied a redshift and flux limit on our lens samples. We used lenses with $z\leq0.5$ and SDSS $r$-band magnitude brighter than $19.8 $ mag. With these limits, we obtained a number density of lenses of $0.282\, \textrm{arcmin}^{-2}$. 

To mitigate possible biases induced by uneven galaxy pair numbers and matter distributions between the 64 lens galaxy maps, we subtracted the lensing signal around random points. This is similar to GGL studies, where the shear around random positions is measured and subtracted from the original measurement \citep{Singh2017}. However, for G3L, this task is in general more difficult because it requires a catalogue of unclustered galaxy pairs. Creating such a catalogue is in general non-trivial because the number of galaxy pairs depends on the selection function of individual galaxies in the survey. Nevertheless, we created a map of unclustered galaxies, similar to GGL, for each simulated lens map by distributing the same number of galaxies as on the lens map at random points on a $4\times 4 \, \textrm{deg}^2$ area. We measured the correlation function $\tilde{\mathcal{G}}$ for each of these random maps and subtracted it from the correlation function measured for the actual lens map.

The total number of triplets to consider for our measurement of $\tilde{\mathcal{G}}_Z$ and $\tilde{\mathcal{G}}_{\rm phys}$ is $5 \times 10^{12}$. This makes the evaluation of the sums in Eq.~\eqref{eq:Gtilde_est_redsh  iftweighted} and Eq.~\eqref{eq:3ptcorrelation_estimator_physical} computationally involved. Because of this computational complexity, third-order correlation functions are usually computed involving some approximation, such as kd-Tree codes \citep{Simon2013}, where galaxy triplets with similar $\vartheta_1, \vartheta_2$ and $\phi$ are averaged. However, we implemented the estimator brute force and calculated it with graphics processing units (GPUs). This approach has two advantages compared to the usual methods. First, if is exact, even at the smallest scales. Second, due to the highly parallelized execution on a GPU, which allows for several thousand simultaneous calculations, the computing time is drastically reduced.In our case, the computational time to process the MR decreased from 200 hours with a kd-Tree code executed on 8 CPU cores to just 9 hours with the brute-force code on a single GPU. Details for our computational implementation are given in Appendix~\ref{app:computation}.

The covariance matrices of the measured $\expval{\mathcal{N}^2 M_\textrm{ap}}$ and $\expval{\mathcal{N}^2 M_\textrm{ap}}_\textrm{phys}$ were computed with jackknifing. For this, we assumed that each of the 64 fields is an independent realization and combined these fields to a total $\expval{\mathcal{N}^2 M_\textrm{ap}}(\theta)$ and 64 jackknife samples $\expval{\mathcal{N}^2 M_\textrm{ap}}_k(\theta),$ where all but the $k$th tile were combined. The covariance matrix is then
\begin{align}
\label{eq:covariancematrix}
C(\theta_i, \theta_j)= \frac{64}{64-1}\sum_{k=1}^{64}& \left[\expval{\mathcal{N}^2 M_\textrm{ap}}_k(\theta_i)-\overline{\expval{\mathcal{N}^2 M_\textrm{ap}}_k}(\theta_i)\right]\\
&\notag \times \left[\expval{\mathcal{N}^2 M_\textrm{ap}}_k(\theta_j)-\overline{\expval{\mathcal{N}^2 M_\textrm{ap}}_k}(\theta_j)\right]\;,
\end{align}
where $\overline{\expval{\mathcal{N}^2 M_\textrm{ap}}_k}(\theta_i)$ is the average of all $\expval{\mathcal{N}^2 M_\textrm{ap}}_k(\theta_i)$. The statistical uncertainty of $\expval{\mathcal{N}^2 M_\textrm{ap}}(\theta_i)$ is $\sigma_i = \sqrt{C(\theta_i, \theta_i)}$. We define the S/N at each scale radius $\theta_i$ as
\begin{equation}
        \textrm{S/N}(\theta_i) = \frac{\expval{\mathcal{N}^2 M_\textrm{ap}}(\theta_i)}{\sigma_i}\;.
\end{equation}

\subsection{Simple mock data}
\label{sec:data:simpleMockData}
Some of our tests also employed simple mock data. These were chosen such that it was easy to create them and to calculate their expected aperture statistics theoretically. For this we used the following assumptions:
\begin{description}
        \item[A.] All matter and galaxies are distributed inside $N_\textrm{h}$ halos over an area $A$.
        \item[B.] All halos are situated on the same lens plane.
        \item[C.] All halos have the same axisymmetric convergence profile $\kappa(\va*{\vartheta})= K\,u(\vartheta)$, where $\int \dd{\vartheta}\, \vartheta\, u(\vartheta) = 1$, and also the same number of galaxies $N_\textrm{gal}$.
        \item[D.] There is no galaxy bias, so the discrete galaxy distribution follows the matter distribution up to Poisson shot-noise.
        \item[E.] Halo centres are distributed randomly within $A$. 
\end{description}

With these assumptions and the calculations in App. \ref{app:calculation}, the theoretical expectation for the aperture statistics using the exponential filter function in Eq.~\eqref{eq:exponentialFilterFunction} is
\begin{align}
\label{eq:NNMapTheo}
&\notag\expval{\mathcal{N}^2 M_\textrm{ap}}(\theta_1, \theta_2, \theta_3)\\
&=\frac{2\pi\,A\,K}{N_\textrm{h}}\int_0^{\infty}\dd{\vartheta}\, \vartheta\, \prod_{i=1}^3 \int_0^{\infty} \dd{y_i}\, \dfrac{y_i\, u(y_i)}{\theta_i^2}\, \exp\left[-\dfrac{(y_i-\vartheta)^2}{2\theta_i^2}\right]\\
&\notag \quad \times \left[\left(1-\frac{y_i^2+\vartheta^2}{2\theta_i^2}\right)\,f_0\left(\dfrac{y_i\,\vartheta}{\theta_i^2}\right) +\dfrac{y_i\vartheta}{\theta_i^2}\,f_1\left(\dfrac{y_i\,\vartheta}{\theta_i^2}\right)\right]\; ,
\end{align}
with $f_n(x) = I_n(x)\, \textrm{e}^{-x}$ and the modified Bessel functions of the first kind $I_n(x)$. We evaluated the integrals numerically with a Monte Carlo integration using the \verb|monte-vegas|-routine of the GNU Scientific Library \citep{GSL}.

We used the Brainerd-Blandford-Smail (BBS) profile \citep{Brainerd1996} as halo convergence profile $\kappa(\va*{\vartheta})$, which is
\begin{equation}
\kappa(\va*{\vartheta}) = \frac{K}{2\pi\,\vartheta\, \theta_\textrm{s}}\, \left(1-\frac{\vartheta}{\sqrt{\vartheta^2 + \theta_\textrm{s}^2}}\right)\,.
\end{equation}
The BBS profile corresponds to a singular isothermal sphere (SIS) for $\vartheta$ much smaller than the scale radius $\theta_s$ that smoothly drops outside the sphere. In contrast to the SIS profile, it has a finite total mass. We chose $K=1\, \text{arcmin}^{2}$ and $\theta_\textrm{s}=\ang{;5;}$.

We created mock lens galaxies following assumptions A to E in a circular area with a radius of $\ang{;700;}$. The lens galaxies were distributed in $2170$ halos with $200$ galaxies each. These numbers were chosen such that the average number density of lens galaxies was $\overline{N}_\textrm{d} = 0.287\, \textrm{arcmin}^{-2}$, the lens number density in our lens sample from the MR. We distributed $3\times 10^6$ source galaxies, whose shear was computed from the halo convergence profiles, in the central $750 \times 750 \, \textrm{arcmin}^2$ area. We only considered lens-lens-source triplets in this area to ensure that the shear of each source was affeted by halos from all directions. No shape noise was added to the shears because our aim was not to create a realistic simulation, but only a simple test case. Because $\tilde{\mathcal{G}}$ is linear in the ellipticities, any shape noise would not bias its estimate and only lead to a larger uncertainty of the measurement. The central area was cut into quadratic tiles with a side length of $\ang{;150;}$, so that finally 25 maps of source and lens galaxies were used.


\section{Results}
\label{sec:results}

\subsection{Effect of the new binning scheme}
\label{sec:results:binning}
\begin{figure}
\resizebox{\hsize}{!}{\includegraphics[width=\linewidth]{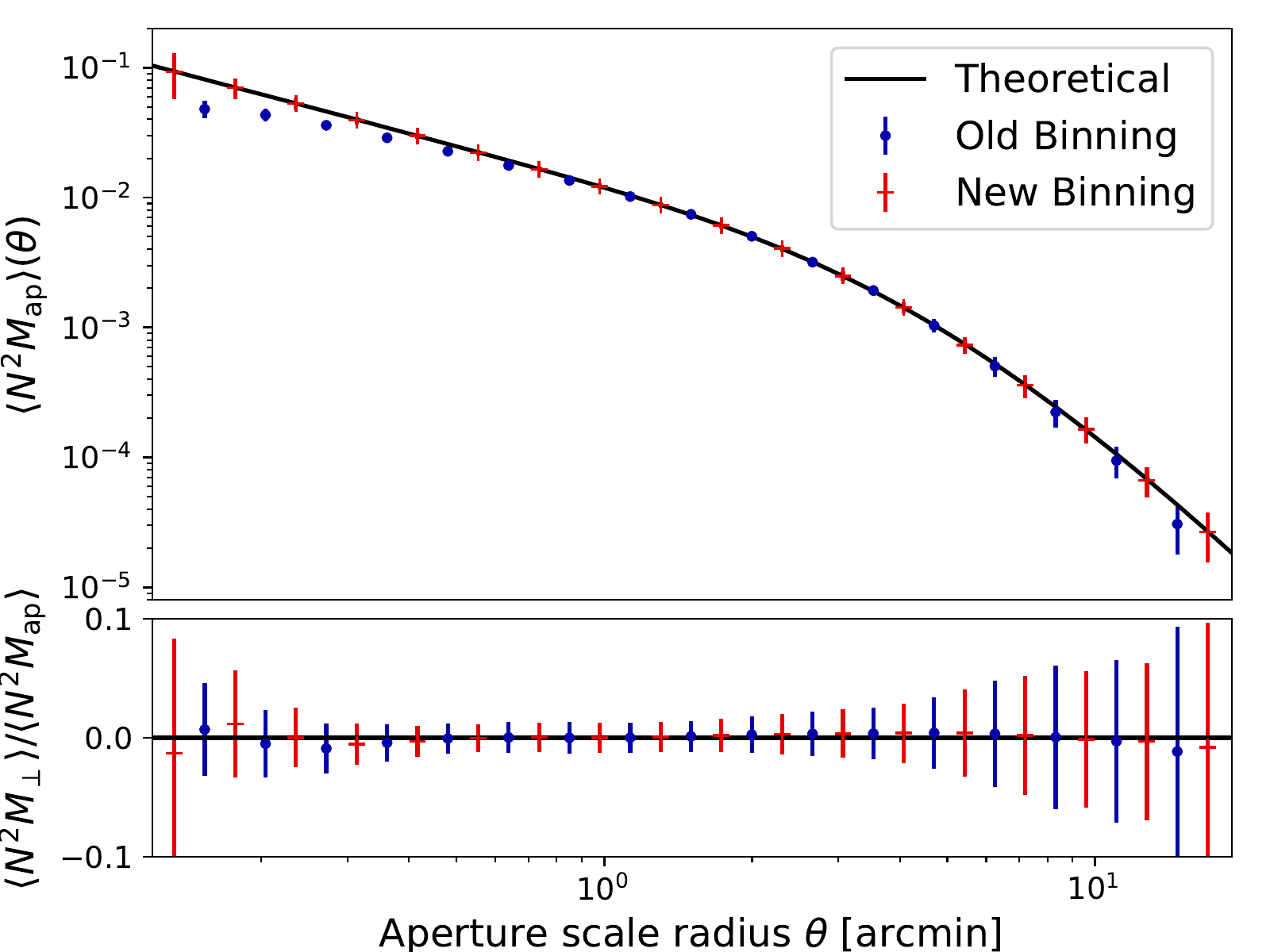}}
\caption{Aperture statistics measured in the simple mock data with the old binning scheme (blue dots) and the new binning scheme (red crosses), as well as the theoretical expectation (black line) given by Eq.~\eqref{eq:NNMapTheo}. The upper plot shows the E mode $\expval{\mathcal{N}^2 M_\textrm{ap}}$ and the lower plot shows the ratio of the B mode $\expval{\mathcal{N}^2 M_\perp}$ and the E mode. Uncertainties are the statistical error estimated with jackknifing.}
\label{fig:Aperture Statistics_mockdata}
\end{figure}

Before measuring the aperture statistics in the data based on the MR, we estimated the effect of the new binning scheme by measuring the aperture statistics for equal-scale radii $\theta$ in the simple mock data, described in Sect~\ref{sec:data:simpleMockData}. The aperture statistics measured in this mock data are displayed in Fig.~\ref{fig:Aperture Statistics_mockdata}. The theoretically expected $\expval{\mathcal{N}^2 M_\textrm{ap}}$ follows a power law for scale radii above $\ang{;2;}$ and steepens for larger scales. The $\expval{\mathcal{N}^2 M_\textrm{ap}}$ measured with the old and the new binning scheme both show the same steepening for $\theta$ larger than $\ang{;2;}$, but the slope of the $\expval{\mathcal{N}^2 M_\textrm{ap}}$ measured with the old binning scheme is considerably shallower for scales between $\ang[astroang]{;0.1;}$ and $\ang[astroang]{;0.6;}$ than the one measured with the new binning scheme. We confirm that the measurement with the new binning scheme agrees with the theoretical expectation within its statistical uncertainty. This agreement of the measured aperture statistics with the theoretical prediction validates our code for estimating $\tilde{\mathcal{G}}$ and for converting $\tilde{\mathcal{G}}$ to $\expval{\mathcal{N}^2 M_\textrm{ap}}$.

\begin{figure}
\resizebox{\hsize}{!}{\includegraphics[width=\linewidth]{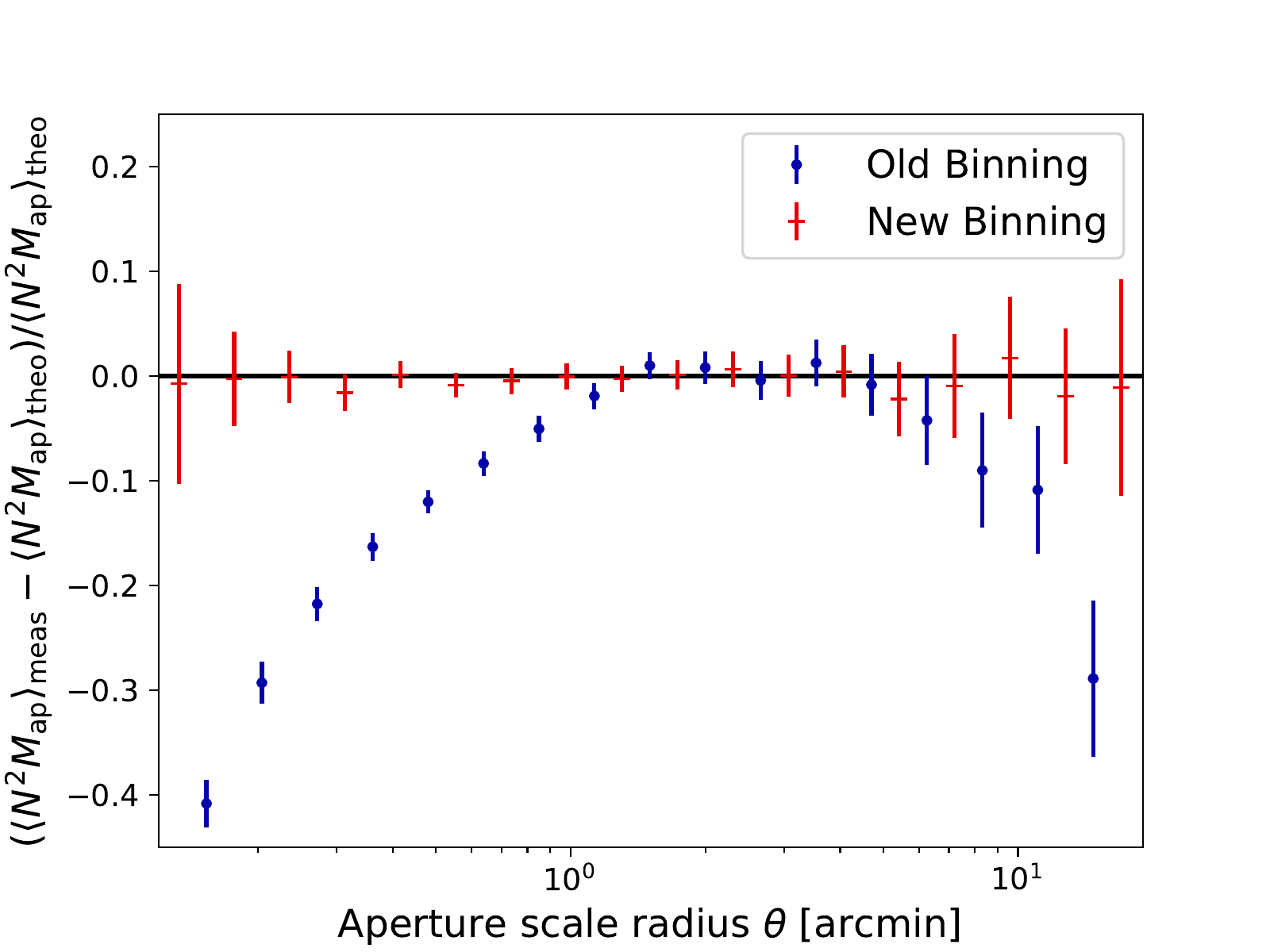}}
\caption{Fractional difference of the measured $\expval{\mathcal{N}^2 M_\textrm{ap}}$ relative to the theoretical prediction in the mock data. Blue dots show the measurement with the old binning scheme; red crosses show the measurement with the new binning scheme.}
\label{fig:bias_mockdata}
\end{figure}

To quantify the effect of the new binning scheme, Fig.~\ref{fig:bias_mockdata} shows the difference of the measured $\expval{\mathcal{N}^2 M_\textrm{ap}}$ to the theoretical prediction for both binning schemes, normalized by the theoretical prediction. While the $\expval{\mathcal{N}^2 M_\textrm{ap}}$ from the old method has no bias at scales between $\ang{;1;}$ and $\ang{;5;}$, it underestimates $\expval{\mathcal{N}^2 M_\textrm{ap}}$ both above and below these scales. At large scales, this bias grows to \SI{10}{\percent} at $\theta=\ang{;10;}$, whereas at small scales, the bias increases with decreasing scale to \SI{40}{\percent} at $\theta=\ang[astroang]{;0.1;}$. The new binning scheme does not show this behaviour. Instead, the bias of the $\expval{\mathcal{N}^2 M_\textrm{ap}}$ measured with the new method is consistent with zero at all considered scales.

\subsection{Effect of lens magnification}

As outlined in Sect.~\ref{sec:methods:magnification bias}, the redshift weighting enables us to measure the impact of lens magnification on G3L. We estimated this effect in the data based on the MR with the two different methods outlined in Sect~\ref{sec:methods:magnification bias}. In the following, the terms ``first'' and ``second'' lens plane refer to the redshift plane in which the lens galaxy lies closer to the observer and closer to the source, respectively.

\begin{figure}
\resizebox{\hsize}{!}{\includegraphics[width=\linewidth]{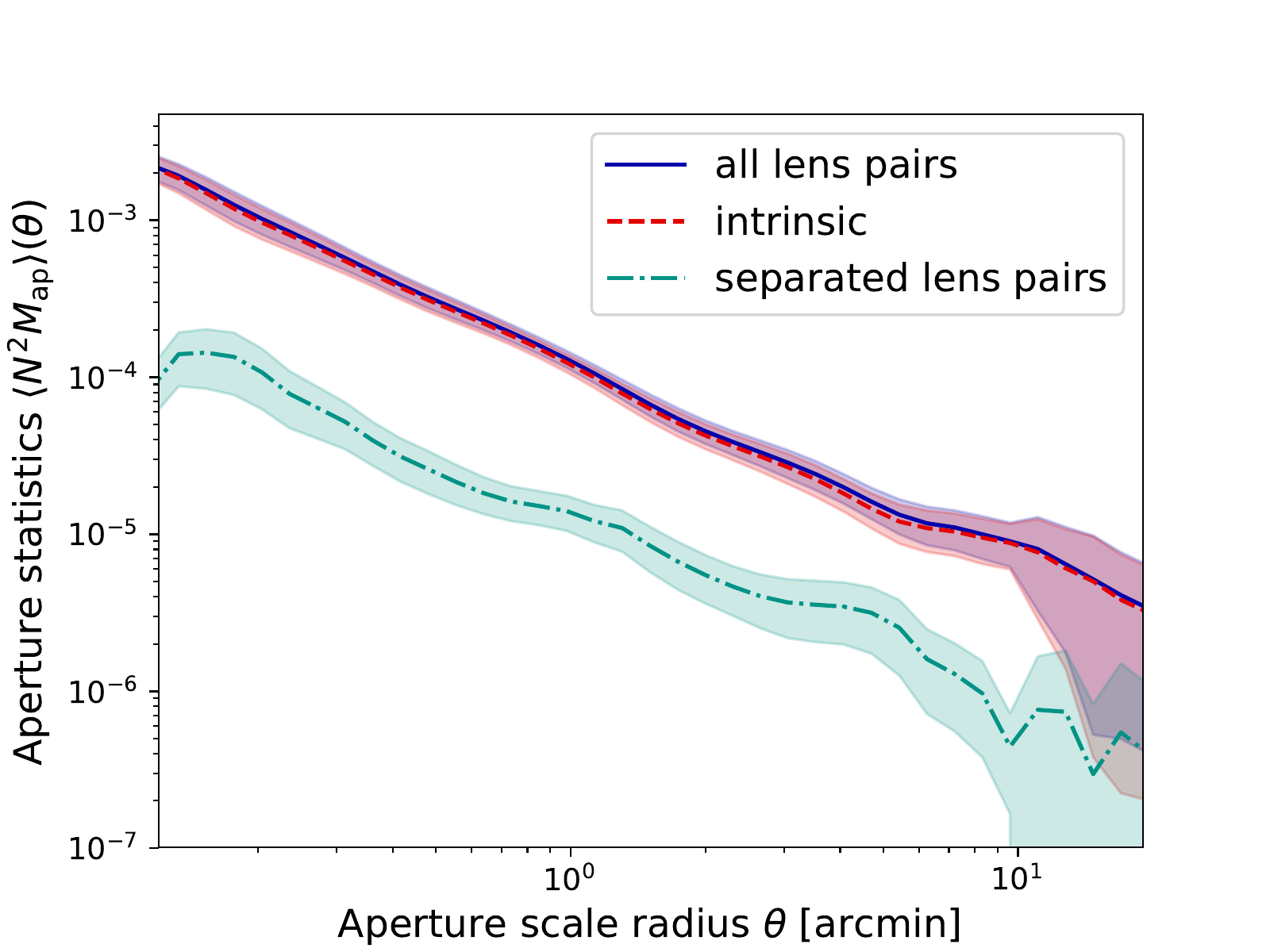}}
\caption{Effect of lens magnification on aperture statistics in the data based on the MR. The green dash-dotted line shows the signal measured for lens pairs with redshift differences larger than $0.01$, which corresponds to the magnification terms in Eq.~\eqref{eq:NNMapWithMagBias}. The blue solid line is the aperture statistics for all lens pairs. The red dashed line is the intrinsic signal, which is corrected for lens magnification by subtracting the signal of separated lens pairs. Shaded regions are the $1\,\sigma$ uncertainties from jackknifing.}
\label{fig:MagBias_MR_angular}
\end{figure}

\begin{figure}
\resizebox{\hsize}{!}{\includegraphics[width=\linewidth]{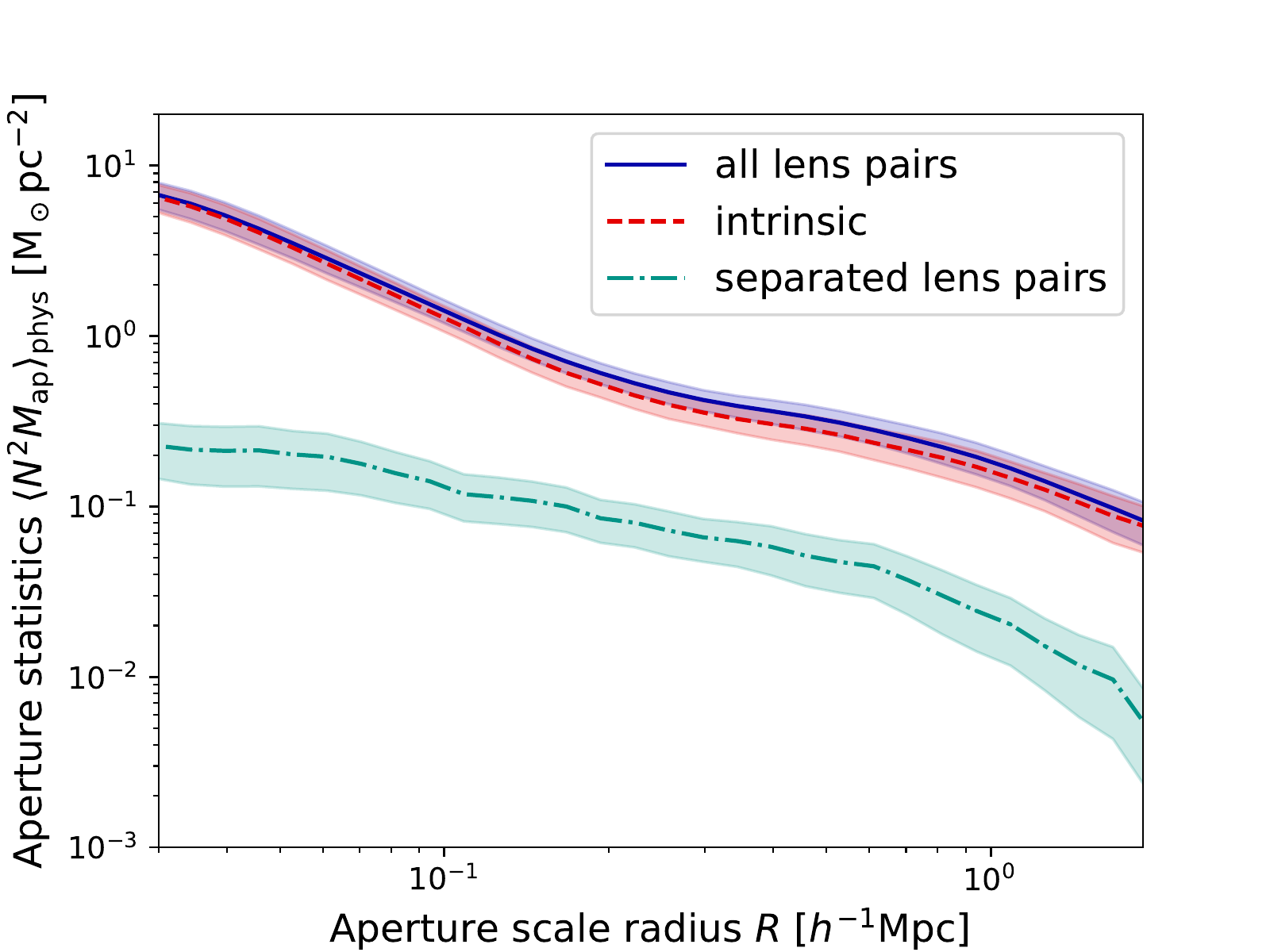}}
\caption{Same as Fig.~\ref{fig:MagBias_MR_angular}, but for physical aperture statistics.}
\label{fig:MagBias_MR_physical}
\end{figure}

The result for the first method, using the step function weighting in the estimation of $\tilde{\mathcal{G}}_Z$, is shown in Figs.~\ref{fig:MagBias_MR_angular} and \ref{fig:MagBias_MR_physical} for $\expval{\mathcal{N}^2 M_\textrm{ap}}$ and $\expval{\mathcal{N}^2 M_\textrm{ap}}_{\rm phys}$, respectively. The figures show the aperture statistics measured for lens pairs with redshift differences larger than $0.01$. If there were no lens magnification, this signal should vanish. The figures also show the aperture statistics measured when all lens pairs are taken into account, as well as the intrinsic aperture statistics ``corrected'' for the effect of lens magnification by subtracting the signal of physically distant lens pairs from the total measured aperture statistics.

For both $\expval{\mathcal{N}^2 M_\textrm{ap}}$ and $\expval{\mathcal{N}^2 M_\textrm{ap}}_{\rm phys}$, the signal of physically separated lens pairs is non-zero. We attribute this signal to the three magnification terms in Eq.~\eqref{eq:NNMapWithMagBias}. For $\expval{\mathcal{N}^2 M_\textrm{ap}}$ this signal is approximately \SI{10}{\percent} of the signal of all lens pairs. For $\expval{\mathcal{N}^2 M_\textrm{ap}}_{\rm phys}$, the magnification leads to a slightly weaker additional signal at scales below $0.1 h^{-1}$ Mpc and approximately \SI{10}{\percent} at larger scales. 

At angular scales smaller than $\ang[astroang]{;0.2;}$, the signal due to lens magnification for $\expval{\mathcal{N}^2 M_\textrm{ap}}$ decreases. This is probably due to smoothing in the simulation, which is no longer accurate at these small angular scales. Smoothing flattens the centre of halo convergence profiles in the simulation. If the aperture statistics are measured at scale radii smaller than the smoothing lengths, the flattened profile then leads to a smaller measured signal.

\begin{figure}
        \resizebox{\hsize}{!}{\includegraphics[width=\linewidth]{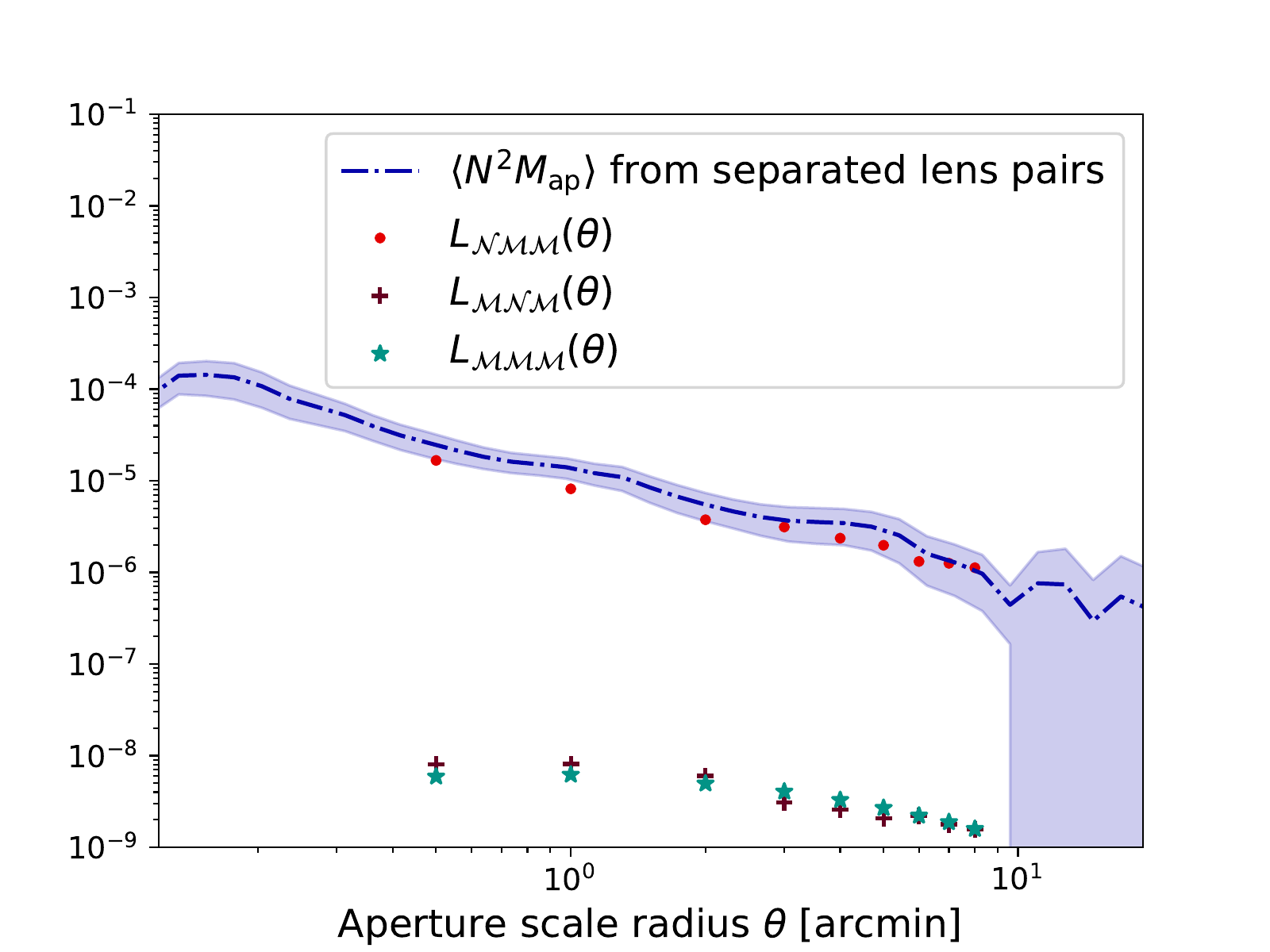}}
        \caption{Individual lens magnification terms in the MR. Green stars depict the term due to correlation between the convergence maps at the two lens planes and at the source plane, red dots are the term due to correlation of the galaxy number density at the first lens plane to the convergence maps at the second lens and the source plane, brown crosses are the term due to correlation of the convergence maps at the first lens and the source plane to the galaxy number density at the second lens plane. The blue line is the measured $\expval{\mathcal{N}^2 M_\textrm{ap}}$ for separated lens pairs, which should correspond to the total lens magnification signal.}
        \label{fig:MagBias_indTerms}
\end{figure}

To verify that the measured signal for distant lens pairs is indeed related to lens magnification, Fig.~\ref{fig:MagBias_indTerms} shows the magnification terms measured with the second method from Sect.~\ref{sec:methods:magnification bias}, using the convolution of the aperture filter function with the convergence and number density maps. For comparison, the measured $\expval{\mathcal{N}^2 M_\textrm{ap}}$ from the first method is also shown.

The figure shows that $L_\mathcal{NMM}(\theta)$, which is due to the correlation of the number density of galaxies at smaller redshift to the convergence measured at higher redshifts, is the dominating term. It is larger than $L_\mathcal{MMM}$ and $L_\mathcal{MNM}$ by three orders of magnitude. Furthermore, the correlation of three convergence maps $L_\mathcal{MMM}$ and the correlation of foreground convergence maps to background galaxies $L_\mathcal{MNM}$ are almost identical. Consequently, the total lens magnification signal is approximately $L_\mathcal{NMM}$. 

This indicates that the lens magnification signal is driven mainly by the correlation of matter and the galaxy distribution at the first lens plane. This matter affects the convergence at the second lens and the source plane and thereby causes a significant $L_\mathcal{NMM}$. Neither $L_\mathcal{MNM}$ nor $L_\mathcal{MMM}$ depend on the correlation between matter and galaxies at the same plane, and they are mainly caused by the LSS in front of the first lens plane. This LSS influences $\mathcal{N}_\theta$ and $M_{\textrm{ap}, \theta}$ at the lens planes and the source plane and thereby induces the non-zero $L_\mathcal{MNM}$ and $L_\mathcal{MMM}$. However, as shown in Fig.~\ref{fig:MagBias_indTerms}, this effect is secondary, and the LSS in front of the lenses does not have a strong effect on the overall signal.

The total lens magnification signal is of the same order of magnitude as the $\expval{\mathcal{N}^2 M_\textrm{ap}}$ measured with separated lens pairs. At scales above \ang[astroang]{;1;} , it indeed agrees with the measured $\expval{\mathcal{N}^2 M_\textrm{ap}}$ for separated lens pairs within its statistical uncertainty. At smaller scales, the difference between the two quantities is still smaller than twice the statistical uncertainty. According to Eq.~\eqref{eq:NNMapWithMagBias2}, the intrinsic aperture statistics are
\begin{align}
        &\expval{\mathcal{N}_{0,\theta}(\vec{\vartheta}, z_1)\, \mathcal{N}_{0,\theta}(\vec{\vartheta}, z_2)\, M_{\textrm{ap},\theta}(\vec{\vartheta})}\\
        &=\notag \expval{\mathcal{N}^2 M_\textrm{ap}}(\theta) - L_\mathcal{NMM}(\theta) - L_\mathcal{MNM}(\theta) - L_\mathcal{MMM}(\theta)\;,
\end{align}
where $\expval{\mathcal{N}^2 M_\textrm{ap}}$ are the measured aperture statistics for separated lens pairs. Therefore, the intrinsic aperture statistics for separated lens pairs vanishes, as expected.

\subsection{Effect of redshift weighting}

\begin{figure*}
        \centering
        \begin{subfigure}[b]{0.5\linewidth}
                \includegraphics[width=\linewidth]{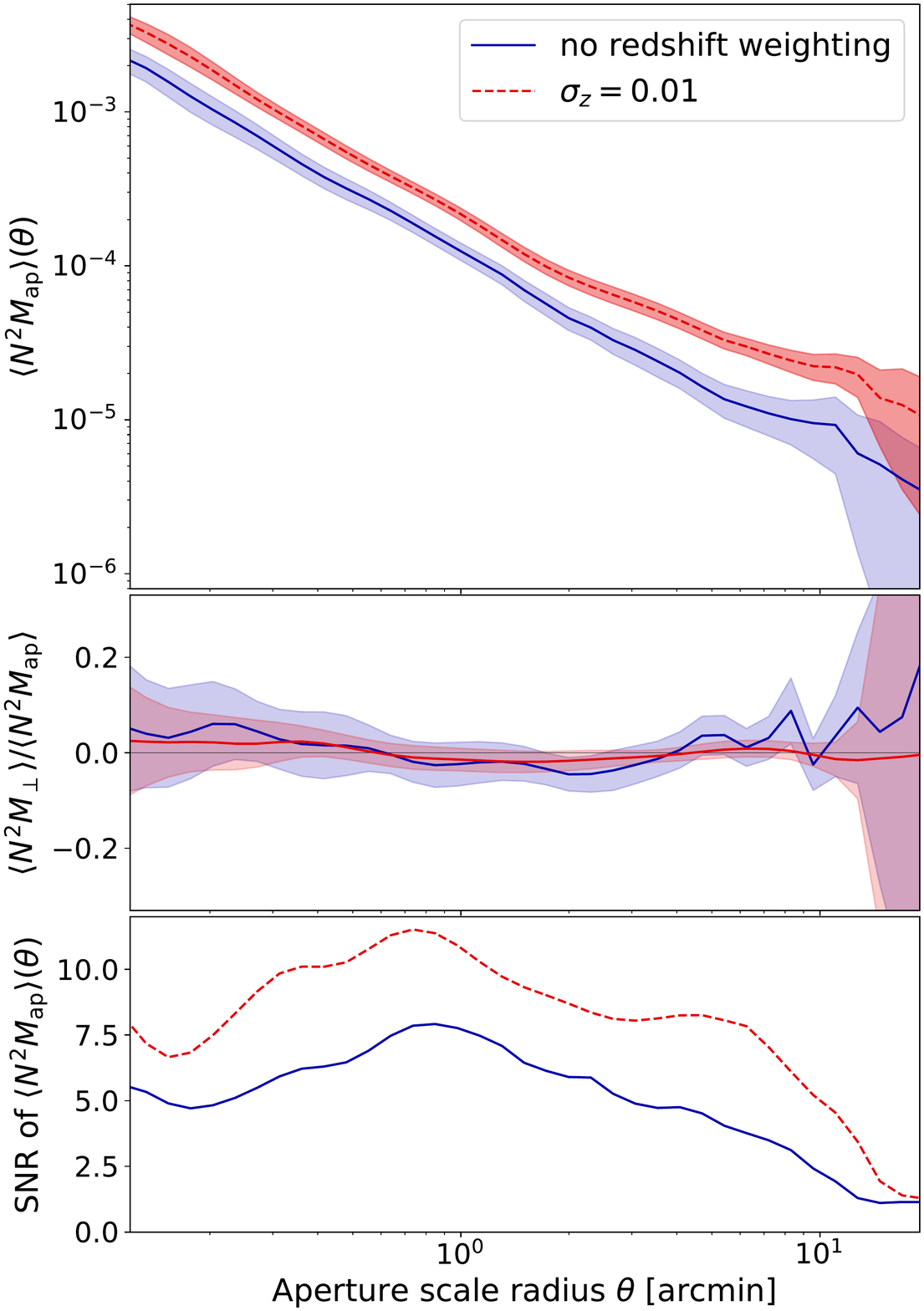}
                \caption{Aperture statistics in angular units}
                \label{fig:NNMap_angular}
        \end{subfigure}%
        \begin{subfigure}[b]{0.5\linewidth}
                \includegraphics[width=\linewidth]{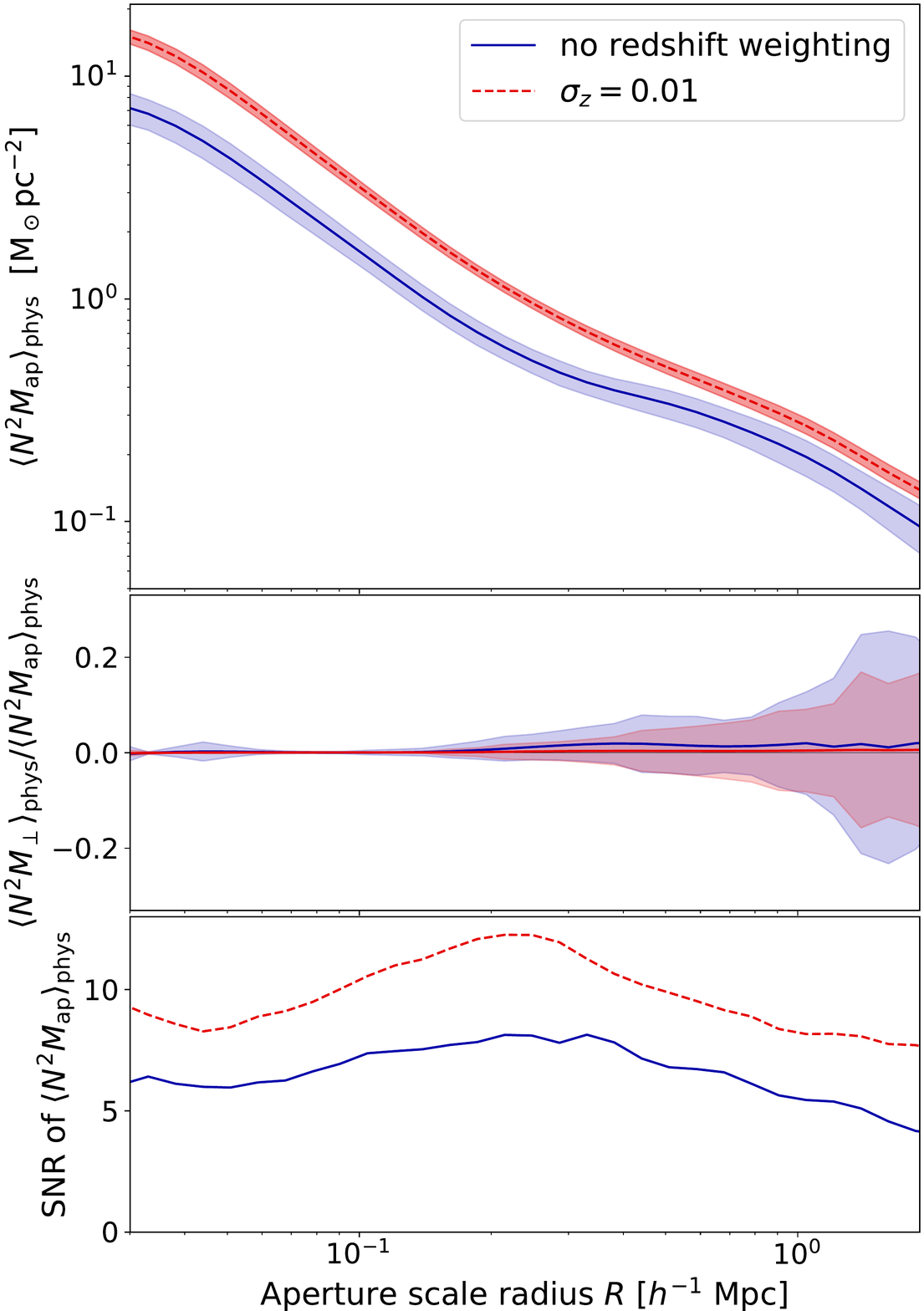}   
                \caption{Aperture statistics in physical units}
                \label{fig:NNMap_phys}
        \end{subfigure}
\caption{Aperture statistics measured in the mock data based on the MR, in (\subref{fig:NNMap_angular}) with angular and in (\subref{fig:NNMap_phys}) with physical units. The upper plots show the E modes $\expval{\mathcal{N}^2 M_\textrm{ap}}$ and $\expval{\mathcal{N}^2 M_\textrm{ap}}_\textrm{phys}$, and the middle plot shows the ratio of the B modes $\expval{\mathcal{N}^2 M_\perp}$ and $\expval{\mathcal{N}^2 M_\perp}_\textrm{phys}$ to $\expval{\mathcal{N}^2 M_\textrm{ap}}$ and $\expval{\mathcal{N}^2 M_\textrm{ap}}_\textrm{phys}$. The lower plots give the S/N of $\expval{\mathcal{N}^2 M_\textrm{ap}}$ and $\expval{\mathcal{N}^2 M_\textrm{ap}}_\textrm{phys}$. The red dashed line depicts $\expval{\mathcal{N}^2 M_\textrm{ap}}$ taken with a redshift weighting function with width $\sigma_z=0.01$. Shaded regions show the $1\sigma$ uncertainties from jackknifing. The blue solid line denotes the measurements without redshift weighting.}
\label{fig:NNMap}
\end{figure*}

The results for $\expval{\mathcal{N}^2 M_\textrm{ap}}$ for the data based on the MR with and without redshift weighting are shown in Fig.~\ref{fig:NNMap_angular}. The measured $\expval{\mathcal{N}^2 M_\perp}$ is consistent with zero, both with and without redshift weighting. This signifies that no indication of parity violation and B-modes is found in the simulation. 

Redshift-weighting increases the S/N, as indicated by the decreasing error region in Fig.~\ref{fig:NNMap_angular}. Simultaneously, the measured $\expval{\mathcal{N}^2 M_\textrm{ap}}$ is increased by a factor of approximately two. This is expected because redshift weighting is assumed to increase both signal and S/N, as discussed in Sect.~\ref{sec:methods:redshift weighting}. The lower plot in Fig.~\ref{fig:NNMap_angular} shows the S/N of $\expval{\mathcal{N}^2 M_\textrm{ap}}$ with and without redshift weighting as function of the aperture scale radius $\theta$. Redshift weighting increases the S/N on all scales. On average, the S/N of $\expval{\mathcal{N}^2 M_\textrm{ap}}$ with redshift weighting is 1.35 times the S/N of $\expval{\mathcal{N}^2 M_\textrm{ap}}$ without redshift weighting.  

The measured physical aperture statistics $\expval{\mathcal{N}^2 M_\textrm{ap}}_\textrm{phys}$ are displayed in Fig.~\ref{fig:NNMap_phys}. Again, the B mode is consistent with zero at all scales. Redshift weighting increases the signal by a factor of two, similar to the increase of $\expval{\mathcal{N}^2 M_\textrm{ap}}$, whereas the error region decreases. The increase of the S/N of $\expval{\mathcal{N}^2 M_\textrm{ap}}_{\rm phys}$, shown in the lower plot of Fig.~\ref{fig:NNMap_phys}, is at the same level as for $\expval{\mathcal{N}^2 M_\textrm{ap}}$; on average, the S/N increases by \SI{34}{\percent}. In Fig.~\ref{fig:NNMap} the S/N of $\expval{\mathcal{N}^2 M_\textrm{ap}}_{\rm phys}$ is higher than the S/N of $\expval{\mathcal{N}^2 M_\textrm{ap}}$, both with and without redshift weighting.


\section{Discussion}
\label{sec:discussion}

We proposed three improvements to the measurement of the G3L signal: Using a redshift weighting of lens galaxies to improve the precision, removing biases on the estimator with a new binning scheme, and accounting for the impact of lens magnification. We furthermore showed how the G3L signal can be measured in physical units.

The effect of the improved binning scheme can be seen by comparing our measurement on the simple mock data with the theoretical expectation. The original binning leads to a discrepancy between the theoretical expectation and the measurement both for aperture scale radii below $\ang{;1;}$ and above $\ang{;5;}$, whereas the aperture statistics measured with the improved binning agrees with the expectation at all scales. At $\theta=\ang[astroang]{;0.1;}$, the original binning underestimates $\expval{\mathcal{N}^2 M_\textrm{ap}}$ by \SI{40}{\percent}, whereas the result of the new binning scheme agrees with the theoretical expectation. Thus, our new method extends the reliability of the measurement. This is achieved by the tessellation because now the three-point correlation function is not incorrectly set to zero in bins for which no lens-lens-source triplet is found.

The signal due to the magnification of lens galaxies is approximately \SI{10}{\percent} of the total G3L signal and can therefore not be neglected in theoretical modelling of the G3L signal. Previous studies \citep{Simon2008, Simon2013} did not account for lens magnification. Nonetheless, even though it has a significant effect on the measured G3L signal, the conclusions of \citet{Saghiha2017}, who found good agreement between the G3L measured in CFHTLenS and the MR with the SAM by \citetalias{Henriques2015}, are not impaired by this because the observational data and the simulations both included lens magnification.

We also demonstrated how the effect of lens magnification can be corrected for. The additional signal due to this effect can be measured with our redshift weighting by considering only lens pairs that are sufficiently far separated along the line of sight. The resulting signal matches the expectation for lens magnification from the convergence and number density maps at different redshift slices. We therefore conclude that the lens magnification signal can indeed be measured  using physically separated lens pairs that have no intrinsic correlation. Because lens magnification only causes an additive signal, the intrinsic correlation can be found by subtracting the additional component from the overall measurement.

By directly measuring the different terms due to lens magnification, we found that the dominating term because galaxies at the closer lens plane are correlated to the convergence measured at the second lens plane and the convergence measured at the source plane, whereas the other terms are three orders of magnitude smaller. This finding explains why we measure a significant signal due to lens magnification, even though previous studies \citep[e.g.][]{Simon2013} expected this effect to be negligible: In these evaluations, only the $M_\textrm{ap}^3$ term was considered, which is indeed much smaller than any $\expval{\mathcal{N}^2 M_\textrm{ap}}$ signal. However, as we have shown here, it is not the dominant term for lens magnification.

The magnification signal is mainly due to correlation of galaxies with matter at the first lens plane, which influences the convergence at the second lens and the source plane. Matter in front of both lenses, which influences the observed lens number density and the convergence at both lens and the source plane, also contributes to the magnification signal, although its measured contribution is minor. Nonetheless, because our lens sample has a low median redshift of $0.2$, the effect of foreground matter might be stronger for lens samples at higher redshifts.

Using redshift weighting, we increased the S/N of both $\expval{\mathcal{N}^2 M_\textrm{ap}}$ and $\expval{\mathcal{N}^2 M_\textrm{ap}}_{\rm phys}$ by approximately \SI{35}{\percent} between $\ang[astroang]{;0.1;}$ and $\ang[astroang]{;10;}$ and $0.1 h^{-1} \textrm{Mpc}$ and $2 h^{-1} \textrm{Mpc}$. Simultaneously, the signal was increased by a factor of approximately two. This meets our expectation that the signal increases by the square of the increase in S/N.

Our choice of $\sigma_z$ was motivated by the correlation length between galaxies, the redshift distribution of galaxy pairs, and the typical peculiar velocities of galaxies in clusters. Choosing a different $\sigma_z$ will lead to a different measured signal and to a different increase in S/N. However, the choice of $\sigma_z$ does not affect the physical interpretation of the aperture statistics as long as the same $\sigma_z$ is chosen in the theoretical modelling. Moreover, for each survey, different values of $\sigma_z$ can be chosen, and the value that provides the highest S/N increase can be retained.

For the redshift-weighting scheme in the MR, we could use exact redshifts for all lens galaxies. This is generally not possible for observations. Although the redshift weighting with a broad weighting function might be possible for lens galaxies with photometric redshift estimates, we expect that redshift weighting is most useful for data sets that include spectroscopic redshifts. The uncertainties of spectroscopic redshifts are much smaller than of those photometric redshifts, so that a narrow weighting function, such as the one chosen for this work, can be used.

At first glance, the measurement of the aperture statistics in physical units $\expval{\mathcal{N}^2 M_\textrm{ap}}_{\rm phys}$ does not appear to provide additional information to the measurement in angular units. However, in contrast to $\expval{\mathcal{N}^2 M_\textrm{ap}}$, $\expval{\mathcal{N}^2 M_\textrm{ap}}_{\rm phys}$ is independent of the source redshift distribution. Direct comparisons of $\expval{\mathcal{N}^2 M_\textrm{ap}}_{\rm phys}$ between surveys with different galaxy distributions are possible. Furthermore, the S/N of $\expval{\mathcal{N}^2 M_\textrm{ap}}_{\rm phys}$ is slightly higher than for $\expval{\mathcal{N}^2 M_\textrm{ap}}$, independent of the redshift weighting. This is because for $\expval{\mathcal{N}^2 M_\textrm{ap}}_{\rm phys}$, triplets are weighted according to their lensing efficiency.

We only applied our improvements on the lens-lens-shear correlation function and the aperture statistics $\expval{\mathcal{N}^2 M_\textrm{ap}}$ here. However, the new binning scheme can also be applied to measurements of the lens-shear-shear correlation and $\expval{\mathcal{N} M_\textrm{ap}^2}$. We expect that this might extend the accuracy of measurement of this aperture statistics to scales below $\ang{;1;}$, which were not taken into account in previous measurements \citep{Simon2013}. The transformation into physical units can also be applied to $\expval{\mathcal{N} M_\textrm{ap}^2}$.

\begin{acknowledgements}
        We are grateful to the anonymous referee for providing helpful comments. We thank Sandra Unruh for providing code for the slopes of the galaxy luminosity function in the MR.
LL is a member of and received financial support for this research from the International Max Planck Research School (IMPRS) for Astronomy and Astrophysics at the Universities of Bonn and Cologne.
\end{acknowledgements}

\bibliographystyle{aa} 
\bibliography{biblio} 


\begin{appendix}
\section{Calculation of aperture statistics for mock data}
\label{app:calculation}
Averages in the halo model are given by
\begin{align}
\expval{f} 
&=  \int \dd{m_1} \dots \dd{m_{N_\textrm{h}}}\;\underbrace{P_\textrm{m}(m_1, \dots, m_{N_\textrm{h}})}_{\substack{\text{Probability that haloes}\\\text{have masses }m_1, \dots, m_2}}\\
&\quad \times \notag \int \dd[3]{x_1} \dots \dd[3]{x_{N_\textrm{h}}}\; \underbrace{P_\textrm{h}(\va{x}_1, \dots, \va{x}_{N_\textrm{h}}\;|\; m_1, \dots , m_{N_\textrm{h}})}_{\substack{\text{Probability that halo centres are at }\va{x}_1, \dots, \va{x}_{N_\textrm{h}}}} \\
&\quad \times \notag \int \dd[3]{\Delta \va{x}_{11}} \dots \int \dd[3]{\Delta \va{x}_{N_{\textrm{h}}N_{\textrm{gal}}}}\;\\
&\quad \quad \notag \underbrace{P_\textrm{gal}(\Delta \va{x}_{11}, \dots, \Delta \va{x}_{N_{\textrm{h}}N_{\textrm{gal}}}\;|\; \va{x}_1, \dots, \va{x}_{N_\textrm{h}}, m_1, \dots , m_{N_\textrm{h}})}_{\substack{\text{Probability that galaxies are at }\Delta \va{x}_{11}, \dots, \Delta \va{x}_{N_{\textrm{h}}N_{\textrm{gal}}} \\\text{ if the halos are at }\va*{\vartheta}_1, \dots, \va*{\vartheta}_{N_\textrm{h}}}}\, f\;.
\end{align}
Using assumption B in Sect.~\ref{sec:results:binning}, we can reduce this integration to two spatial dimensions and use the projected halo centres $\va*{\vartheta}_i$ and the projected separation $\Delta \va*{\vartheta}_{ij}$ of the $j$th galaxy to the $i$th halo centre instead of $\va{x}_i$ and $\Delta \va{x}_{ij}$. Furthermore, due to assumption C, the mass integrals are trivial. Assumption D leads to 
\begin{align}
&P_\textrm{gal}(\Delta \vartheta_{11}, \dots, \Delta \vartheta_{N_\textrm{h}N_{\textrm{gal}}}\;|\; \va*{\vartheta}_1, \dots, \va*{\vartheta}_{N_\textrm{h}}, m_1, \dots , m_{N_\textrm{h}})\\
&=\notag u(\Delta \vartheta_{11})\dots u(\Delta \vartheta_{N_\textrm{h}N_{\textrm{gal}}})\;.
\end{align}
Assumption E means that
\begin{align}
P_\textrm{h}(\va*{\vartheta}_1, \dots, \va*{\vartheta}_{N_\textrm{h}} ) = 
\begin{cases}
{A^{-N_\textrm{h}}} &\text{for }\,(\va*{\vartheta}_1, \dots, \va*{\vartheta}_{N_\textrm{h}} ) \in A \\
0 &\text{else}
\end{cases}\; ,
\end{align}
so that the average of a quantity is given by
\begin{align}
\expval{f}&= {A^{-N_\textrm{h}}}\, \int_A \dd[2]{\vartheta_1} \dots \dd[2]{\vartheta_{N_\textrm{h}}}\,\\
&\notag \quad \times \int \dd[2]{\Delta \vartheta_{11}}\, \dots \dd[2]{\Delta \vartheta_{N_\textrm{h}N_{\textrm{gal}}}}\, u_1({\Delta \vartheta_{11}})\, \dots \, u_{N_\textrm{h}}({\Delta \vartheta_{N_\textrm{h}N_{\textrm{gal}}}})\,f\;.
\end{align}
Consequently, the correlation function $\expval{N(\va*{\theta}_1)\,N(\va*{\theta}_2)\,\kappa(\va*{\theta}_3)}$ of the galaxy number density $N(\va*{\theta})$ and the projected matter density $\kappa$ is
\begin{align}
\label{eq:NNkappaHaloModel}
&\expval{N(\va*{\theta}_1)\,N(\va*{\theta}_2)\,\kappa(\va*{\theta}_3)}\\
&=\notag {A^{-N_\textrm{h}}}\, \int_A \dd[2]{\vartheta_1} \dots \dd[2]{\vartheta_{N_\textrm{h}}}\, \\
&\notag \quad\times \int \dd[2]{\Delta \vartheta_{11}}\, \dots \dd[2]{\Delta \vartheta_{N_\textrm{h}N_{\textrm{gal}}}}\, u_1({\Delta \vartheta_{11}})\, \dots \, u_{N_\textrm{h}}({\Delta \vartheta_{N_\textrm{h}N_{\textrm{gal}}}})\,\\
&\notag \quad\times N(\va*{\theta}_1)\,N(\va*{\theta}_2)\,\kappa(\va*{\theta}_3)\;.
\end{align}
The matter density $\kappa$ is the sum of the convergence profiles of all halos,
\begin{equation}
\label{eq:kappaHaloModel}
\kappa(\va*{\theta}) = K\, \sum_{i=1}^{N_\textrm{h}} u\left(|\va*{\theta} - \va*{\vartheta}_i|\right)\;.
\end{equation}
We treat galaxies as discrete objects, therefore their number density is
\begin{equation}
\label{eq:NHaloModel}
N(\va*{\theta}) = \sum_{i=1}^{N_\textrm{h}} \sum_{j=1}^{N_{\textrm{gal}}} \delta_\textrm{D}(\va*{\theta} - \va*{\vartheta}_i - \Delta\va*{\vartheta}_{ij})\;.
\end{equation}
Inserting Eq.~\eqref{eq:kappaHaloModel} and Eq.~\eqref{eq:NHaloModel} into Eq.~\eqref{eq:NNkappaHaloModel} leads to
\begin{align}
&\expval{N(\va*{\theta}_1)\,N(\va*{\theta}_2)\,\kappa(\va*{\theta}_3)}\\
&=\notag A^{-N_\textrm{h}}\, K\, \int_A \dd[2]{\vartheta_1} \dots \dd[2]{\vartheta_{N_\textrm{h}}}\, \\
&\quad\notag \times \int \dd[2]{\Delta \vartheta_{11}}\, \dots \dd[2]{\Delta \vartheta_{N_\textrm{h}N_{\textrm{gal}}}}\, u_1({\Delta \vartheta_{11}})\, \dots \, u_{N_\textrm{h}}({\Delta \vartheta_{N_\textrm{h}N_{\textrm{gal}}}})\\
&\quad\notag \times \sum_{i=1}^{N_\textrm{h}} \sum_{j=1}^{N_\textrm{h}} \sum_{k=1}^{N_\textrm{h}} \, u(|\va*{\theta}_1 - \va*{\vartheta}_i|)\, \sum_{l=1}^{N_\textrm{gal}} \delta_\textrm{D}(\va*{\theta}_2 - \va*{\vartheta}_j - \Delta\va*{\vartheta}_{jl}) \, \\
&\quad \notag \times \sum_{m=1}^{N_\textrm{gal}} \delta_\textrm{D}(\va*{\theta}_3 - \va*{\vartheta}_k - \Delta\va*{\vartheta}_{km})\,.
\end{align}
The delta `functions' reduce the integrals, therefore the expression simplifies to
\begin{align}
        &\notag \expval{N(\va*{\theta}_1)\,N(\va*{\theta}_2)\,\kappa(\va*{\theta}_3)}\\
        &=A^{-N_\textrm{h}}\, K\,  \int_A \dd[2]{\vartheta_1} \dots \dd[2]{\vartheta_{N_\textrm{h}}}\,\\
        &\quad\notag\sum_{i=1}^{N_\textrm{h}} \sum_{j=1}^{N_\textrm{h}} \sum_{k=1}^{N_\textrm{h}} \sum_{l=1}^{N_\textrm{gal}} \sum_{m=1}^{N_\textrm{gal}}\, u(|\va*{\theta}_1 - \va*{\vartheta}_i|) \,  u(|\va*{\theta}_2 - \va*{\vartheta}_j|) \, u(|\va*{\theta}_3 - \va*{\vartheta}_k|)  \\
        &=A^{-N_\textrm{h}}\, K\, N_\textrm{gal}^2 \int_A \dd[2]{\vartheta_1} \dots \dd[2]{\vartheta_{N_\textrm{h}}}\,\\
        &\quad \notag \sum_{i=1}^{N_\textrm{h}} \sum_{j=1}^{N_\textrm{h}} \sum_{k=1}^{N_\textrm{h}} \, u(|\va*{\theta}_1 - \va*{\vartheta}_i|) \,  u(|\va*{\theta}_2 - \va*{\vartheta}_j|) \, u(|\va*{\theta}_3 - \va*{\vartheta}_k|).  
\end{align}
We can split this triple sum into a one-halo term with $i=j=k$, three two-halo terms with $i=j\neq k$, $i=k \neq k$ and $j=k\neq i,$ and a three-halo term with $i\neq j \neq k$. When we use $\int_A \dd[2]{\vartheta} = A$ and $\int \dd[2]{\vartheta} u(\vartheta) = 1$, this leads to
\begin{align}
&\notag \expval{N(\va*{\theta}_1)\,N(\va*{\theta}_2)\,\kappa(\va*{\theta}_3)}\\
&= A^{-N_\textrm{h}}\, K\, N_{\textrm{gal}}^2 \sum_{i=1}^{N_\textrm{h}}\, A^{N_\textrm{h}-1}\,\\
&\notag \quad \times \int \dd[2]{\vartheta}\; u\left(|\va*{\theta}_1 - \va*{\vartheta}|\right)\, u\left(|\va*{\theta}_2 - \va*{\vartheta}|\right)\, u\left(|\va*{\theta}_3 - \va*{\vartheta}|\right) \\
&\notag \quad + A^{-N_\textrm{h}}\, K\, N_{\textrm{gal}}^2 \sum_{i=1}^{N_\textrm{h}} \sum_{j\neq i} A^{N_\textrm{h}-2}\,  \int \dd[2]{\vartheta}\; u\left|\va*{\theta}_1 - \va*{\vartheta}|\right)\, u\left(|\va*{\theta}_3 - \va*{\vartheta}|\right)\\
&\notag \quad + A^{-N_\textrm{h}}\, K\, N_{\textrm{gal}}^2 \sum_{i=1}^{N_\textrm{h}} \sum_{j\neq i} A^{N_\textrm{h}-2}\,  \int \dd[2]{\vartheta}\; u\left|\va*{\theta}_1 - \va*{\vartheta}|\right)\, u\left(|\va*{\theta}_2 - \va*{\vartheta}|\right)\\
&\notag \quad + A^{-N_\textrm{h}}\, K\, N_{\textrm{gal}}^2 \sum_{i=1}^{N_\textrm{h}} \sum_{j\neq i} A^{N_\textrm{h}-2}\,  \int \dd[2]{\vartheta}\; u\left|\va*{\theta}_2 - \va*{\vartheta}|\right)\, u\left(|\va*{\theta}_3 - \va*{\vartheta}|\right)\\
&\notag \quad+ A^{-N_\textrm{h}}\, K\, N_{\textrm{gal}}^2 \sum_{i=1}^{N_\textrm{h}} \sum_{j\neq i} \sum_{k\neq i, k\neq j} A^{N_\textrm{h}-3}\;\\
&= \frac{N_\textrm{h}\,K\,N_{\textrm{gal}}^2}{A} \int \dd[2]{\vartheta}\; u\left(|\va*{\vartheta}_1 - \va*{\vartheta}|\right)\, u\left(|\va*{\vartheta}_2 - \va*{\vartheta}|\right)\, u\left(|\va*{\vartheta}_3 - \va*{\vartheta}|\right) \\
&\notag \quad+\frac{N_\textrm{h}\,(N_\textrm{h}-1)\,K\,N_{\textrm{gal}}^2}{A^2}  \int \dd[2]{\vartheta}\; u_i\left|\va*{\vartheta}_1 - \va*{\vartheta}|\right)\, u_i\left(|\va*{\vartheta}_3 - \va*{\vartheta}|\right)\\
&\notag\quad +\frac{N_\textrm{h}\,(N_\textrm{h}-1)\,K\,N_{\textrm{gal}}^2}{A^2} \int \dd[2]{\vartheta}\; u\left(|\va*{\vartheta}_1 - \va*{\vartheta}|\right) \, u\left(|\va*{\vartheta}_2 - \va*{\vartheta}|\right)\\
&\notag\quad +\frac{N_\textrm{h}\,(N_\textrm{h}-1)\,K\,N_{\textrm{gal}}^2}{A^2} \int \dd[2]{\vartheta}\; u\left(|\va*{\vartheta}_2 - \va*{\vartheta}|\right)\, u\left(|\va*{\vartheta}_3 - \va*{\vartheta}|\right)\\
&\notag\quad +\frac{N_\textrm{h}\,(N_\textrm{h}-1)\,(N_\textrm{h}-2)\,K\,N_{\textrm{gal}}^2}{A^3}\;.
\end{align}
From this, we can infer $\expval{\mathcal{N}^2 M_\textrm{ap}}$ with Eq.~\eqref{eq:Definition NNMap}. Because the filter function $U_\theta$ is compensated for, the integrals over constant terms vanish and only the first term in the sum remains. Therefore, with the exponential filter function from Eq.~\eqref{eq:exponentialFilterFunction} and $\overline{N}=N_\textrm{h}\,N_\textrm{gal}/A$ leads to Eq.~\eqref{eq:Definition NNMap},
\begin{align}
&\notag\expval{\mathcal{N}^2 M_\textrm{ap}}(\theta_1,\theta_2,\theta_3)\\
&=\dfrac{A\,K}{N_\textrm{h} (2\pi)^3} \int_0^{\infty}\dd[2]{\vartheta}\prod_{i=1}^3 \dfrac{1}{\theta_i^2}\\
&\quad\notag\times \int \dd[2]{\vartheta}_i u\left(|\va*{\vartheta}_i - \va*{\vartheta}|\right) \left(1-\dfrac{\vartheta_i^2}{2\theta_i^2}\right) \exp\left(-\dfrac{\vartheta_i^2}{2\theta_i^2}\right)\\
&= \dfrac{A\,K}{N_\textrm{h} (2\pi)^2} \int_0^{\infty}\dd[2]{\vartheta} \prod_{i=1}^3 \frac{1}{\theta_i^2} \int_0^{2\pi}\dd{\phi_i}\, \int_0^{\infty}\dd{y_i}\; y_i\, u(y_i)\,\\
&\quad \times\notag \exp\left(-\dfrac{y_i^2+\vartheta^2}{2\theta_i^2}\right) \left(1-\dfrac{y_i^2+\vartheta^2}{2\theta_i^2}-\dfrac{y_i\,\vartheta\,\cos(\phi_i)}{\theta_i^2}\right)\;.
\end{align}
We can now use that
\begin{align}
  &\int_0^{2\pi} \dd{x} \; \cos(x)\, \exp(-a\,\cos(x)) = -2\pi\, I_1(a)\,,\\
  &\int_0^{2\pi} \dd{x} \; \exp(-a\,\cos(x)) = 2\pi\, I_0(a)\;,
\end{align}
with the modified Bessel functions of the first kind $I_n$. We also introduce the scaled Bessel functions $f_n(x) = I_n(x)\exp(-x)$, so that the aperture statistics are finally
\begin{align}
&\expval{\mathcal{N}^2 M_\textrm{ap}}(\theta_1, \theta_2, \theta_3)\\
&=\notag\frac{2\pi\,A\,K}{N_\textrm{h}}\int_0^{\infty}\dd{\vartheta}\; \vartheta\, \prod_{i=1}^3 \int_0^{\infty} \dd{y_i}\; \dfrac{y_i\, u(y_i)}{\theta_i^2}\, \exp\left[-\dfrac{(y_i-\vartheta)^2}{2\theta_i^2}\right]\\
&\quad \times \notag\left[\left(1-\frac{y_i^2+\vartheta^2}{2\theta_i^2}\right)f_0\left(\dfrac{y_i\,\vartheta}{\theta_i^2}\right) +\dfrac{y_i\vartheta}{\theta_i^2}f_1\left(\dfrac{y_i\,\vartheta}{\theta_i^2}\right)\right]\;.
\end{align}

\section{Computational implementation with graphics processing units}
\label{app:computation}
Our estimates of $\tilde{\mathcal{G}}$ are computed by calculating the sums in Eqs.~\eqref{eq:Gtilde_est}, \eqref{eq:Gtilde_est_redsh  iftweighted} and \eqref{eq:3ptcorrelation_estimator_physical} brutecforce on a GPU. Our algorithm (see Algorithm~\ref{alg:Gtilde}) for the estimation of $\tilde{\mathcal{G}}$ works similar to the procedures proposed by \citet{Bard2013} for the calculation of the galaxy two-point correlation and by \citet{CardenasMontes2014} for the calculation of the galaxy two- and three-point function and the shear-shear correlation.
It can be used for {calculating the correlation between lenses from the same and from different samples}.

\begin{algorithm}
\caption{Algorithm for computing $\tilde{\mathcal{G}}$}
\label{alg:Gtilde}
\begin{algorithmic}
\State Read in lens and source galaxy positions and source ellipticities into main memory (RAM)
\State Copy galaxy positions and ellipticities from RAM to the GPU
\State Initialize container for $\tilde{\mathcal{G}}$ with $N_\textrm{bins}$ bins on RAM
\State Initialize container for $\tilde{\mathcal{G}}$ with $N_\textrm{bins}$ bins on GPU
\State Initialize $N_\textrm{th}$ threads on GPU
\While{$i$}
\ForAll{sources $j$ with $j \in [i, i + N_\textrm{th}, i + 2N_\textrm{th}, \dots, N_s]$ }
\ForAll{lenses}
\ForAll{lenses}
\State Get index of $\tilde{\mathcal{G}}$ bin for this galaxy triplet
\State Add contribution of this triplet to $\tilde{\mathcal{G}}$ on GPU
\EndFor
\EndFor 
\EndFor
\EndWhile
\State Copy $\tilde{\mathcal{G}}$ from GPU to RAM
\State Write $\tilde{\mathcal{G}}$ to file
\end{algorithmic}
\end{algorithm}

This algorithm is implemented in CUDA 10 using double floating point precision. For the calculation we used an NVIDIA RTX 2080 Ti GPU, which has CUDA capability 7.5 and therefore enables 46 $\times$ 1024 parallel threads. Data were read from and written to ASCII files on an SSD hard drive, enabling fast data transfer.

\end{appendix}

\end{document}